\documentclass[11pt]{article}
\usepackage{blindtext}
\usepackage{titlesec}
\usepackage{hyperref}
\usepackage{amsmath,amsfonts,amssymb}
\hypersetup{dvips,dvipdfm,colorlinks=true,urlcolor=magenta,filecolor=magenta,linktoc=page,citecolor=red,linkcolor=blue,bookmarks=true}
\usepackage{array}
\usepackage{graphicx,epstopdf}
\usepackage{multirow}
\usepackage{xcolor}
\title{A Revisit to Classical and Quantum aspects of Raychaudhuri equation and possible resolution of Singularity}
\date{ }
\begin{document}
		\author{Subenoy Chakraborty \footnote{schakraborty.math@gmail.com (corresponding author)}~~and~~Madhukrishna Chakraborty\footnote{chakmadhu1997@gmail.com}
		\\Department of Mathematics, Jadavpur University, Kolkata - 700032, India}
	\maketitle
		\begin{abstract}
\large In this review, we provide a concrete overview of the Raychaudhuri equation, Focusing theorem and Convergence conditions in a plethora of backgrounds and discuss the consequences. We also present various classical and quantum approaches suggested in the literature that could potentially mitigate the initial big-bang singularity and the black-hole singularity.
	\end{abstract}
	\tableofcontents
	
	\section{Introduction}
\large Phenomenal detection of gravitational waves \cite{LIGOScientific:2017vwq}, \cite{LIGOScientific:2016aoc} has again proved that Einstein's General Relativity (GR) \cite{Wald:1984rg}, \cite{Weinberg:1972kfs}, \cite{Rao:2023nip} is the most well deserving theory of gravity to describe the physical reality. Nevertheless the appearance of singularity \cite{Konkowski:2004pma}, \cite{Stoica:2012cf} in GR sets limit to this theory. Presence of singularities in a physical theory may be indicative of either some remarkable unknowns of the underlying theory or may hint at the modification in the description of the physical reality. For a recent review on Singularities one may refer to \cite{deHaro:2023lbq}. GR fundamentally confronts to three types of singularity namely the black hole space-like crushing singularities (the Schwarzschild case which is the simplest solution of Einstein's field equations) and two cosmological space-like singularities appearing in finite time, namely the Big-Bang singularity \cite{Robertson:1933zz}, \cite{Lifshitz:1963ps}, \cite{Belinski:2014kba}, \cite{Cornish:1996yg} and the Big Rip singularity \cite{Bouhmadi-Lopez:2014cca}. The singularities like black-hole singularity \cite{Hawking:1971vc}, Big-Bang singularity \cite{Robertson:1933zz}, \cite{Lifshitz:1963ps}, \cite{Belinski:2014kba}, \cite{Cornish:1996yg} and the Big Rip singularity \cite{Bouhmadi-Lopez:2014cca} indicate that the physics is no longer described by the classical gravity theory but some quantum version of gravity is needed. On the other hand, the Big Rip \cite{Bouhmadi-Lopez:2014cca} is a future singularity which appears in the context of GR due to a phantom scalar field needed to describe the dark energy era. The resolution of singularities has remained a puzzle over the decades and attracted interest of the relativists and cosmologists. Many have attempted the possible classical as well quantum resolution of these singularities in various works in GR as well as in Modified gravity theories. But it is always interesting to know the root of any underlying theory, equation or problem. \\ \\It was about a half century back when GR was at its infancy and even the simplest solution i.e, the Schwarzschild solution was incomplete. Although FLRW solutions were slowly gaining popularity yet cosmology was at its budding stage. The then relativists including Einstein himself were worried about the existence of singularity in GR. In the early 1950s, Raychaudhuri addressed this issue of singularity while he was working on the features of electronic energy bands in metals. Fascinated by cosmology he wrote a paper which for the first time brought forward the formal geometric definition of a singularity using the notion of `geodesic incompleteness'. He pointed out that singularity is nothing more than an artifact of the symmetries of the matter distribution and deduced an equation named as Raychaudhuri Equation (RE) \cite{Raychaudhuri:1953yv},\cite{Dadhich:2007pi},\cite{Dadhich:2005qr},\cite{Ehlers:2006aa},\cite{Kar:2008zz},\cite{Kar:2006ms},\cite{Chakraborty:2023rgb}. In 1955, the derivation of the RE was published in \cite{Raychaudhuri:1953yv}. Recently, Penrose got Nobel prize for the seminal singularity theorems by Hawking and Penrose himself \cite{Penrose:1964wq},\cite{Hawking:1970zqf},\cite{Hawking:1973uf}. It is the most interesting thing to note that RE \cite{Raychaudhuri:1953yv},\cite{Dadhich:2007pi},\cite{Dadhich:2005qr},\cite{Ehlers:2006aa},\cite{Kar:2008zz},\cite{Kar:2006ms} is the key ingredient of the singularity theorems.   In 1955, the derivation of the RE was published in \cite{Raychaudhuri:1953yv} is the key ingredient behind the singularity theorems. The proof of the theorems involve the notion of geodesic focusing, energy conditions and trapped surfaces. This RE along with the Focusing theorem (FT) proves the inevitable existence of singularity in GR. Thus RE can be treated as a turning point in Einstein gravity theory provided that Strong Energy Condition (SEC) holds. However, since in other relativistic theories of gravity or extension of GR the field equations are different there may be some possibilities to violate the FT even if SEC holds \cite{Chakraborty:2023ork}, \cite{Chakraborty:2023yyz}, \cite{Choudhury:2021zij},\cite{Capozziello:2013fxa}. This directs us to the possible resolution of singularities in Modified gravity theories like $f(R)$ gravity \cite{Chakraborty:2023ork}, $f(T)$ gravity  \cite{Chakraborty:2023yyz} and scalar tensor theory of gravity \cite{Choudhury:2021zij}. Such implications of RE have been discussed in these modified gravity theories with different backgrounds (both homogeneous and inhomogeneous as well as both isotropic and anisotropic) together with possible resolution of singularities under certain physical assumptions. Another reason of formulating RE in modified gravity theories is that extension of GR is considered as a successful candidate to explain the present era of accelerated expansion other than dark energy. \\ \\The review also puts light on  the role of inhomogeneity and anisotropy in convergence and restates the FT for non-zero curvature considering GR at the outset. Further RE in the presence of torsion has been deduced. The importance of the Raychaudhuri scalar in the study of cosmology (in which Bouncing cosmology is of central importance), black-hole singularity and geometry of traversable wormhole has also been taken care of. The geometry behind the equation, outline of its geometric derivation and the essence of the original 1955 paper fall under this review in a quite lucid manner. Besides dealing with these classically, the review also discusses quantum mechanical approaches \cite{Das:2013oda}, \cite{Ali:2014qla}, \cite{Blanchette:2021vid}, \cite{Burger:2018hpz}, \cite{Blanchette:2020kkk}  towards the resolution of singularities. This includes canonical quantization \cite{Chakraborty:2023voy}, formulation of Bohmian trajectories \cite{Chakraborty:2001za} to eliminate the big-bang singularity and resolution of black-hole singularity using Loop Quantum Gravity (LQG) corrections to RE. The motivation behind choosing the quantum theory of gravity in this context is that according to a general speculation, quantum effects which become dominant in strong gravity regimes may alleviate the singularity problem at the classical level \cite{Thiemann:2007pyv}, \cite{DeWitt:1967yk}, \cite{Hawking:1978jz}. Sometimes quantum corrections to the RE gives rise to repulsive terms on the right hand side of the equation leading to possible avoidance of singularity. Therefore, the motivation behind writing this review lies in the fact that the general readership will have a broader idea of what is happening in literature in the context of Raychaudhuri equation and implications of singularity theorems in extended theories of gravity. The layout of the paper is divided into sections and subsections under it as given in the `Contents' section.
	\section{Geometry behind Raychaudhuri equation}
	\subsection{Flow as congruence of geodesic and  Kinematics of Deformable medium}
\large	The celebrated RE deals with the kinematics of flows \cite{Dasgupta:2008jt}. Given a vector field, flows are the integral curves generated by that vector field. The curves are basically of two types namely, geodesic and non-geodesic. In the following sections and subsections we shall mainly review the results on geodesic congruence for, the geodesics play a pivotal role when it comes to gravity. Although the RE is generally true for any curves (time-like or null), its geodesic version can be deduced from the general form. Thus a flow is identified by a congruence of such curves which may be time-like, null or curves having tangent vectors with a positive definite norm in the Euclidean case. Now we put our attention to study the kinematic quantities associated with such flows and how Prof. Amal Kumar Raychaudhuri's equation is related to them \cite{Horwitz:2021lyc}. In this context, it is to be noted that evolution equation of a kinematic quantities (which characterize the flow) in a given space-time background along the flow is of central importance.\\
	
	Let $\tau$ be the parameter labeling points on the curves and $v^{a}$ be the velocity vector field and $R_{ab}$ is the Ricci tensor along the flow. Thus to characterize the flow, one must have various functions in terms of $\tau$. To define the kinematics of a deformable medium, we consider the gradient of the velocity vector field, which can geometrically be represented by a $(0,2)$ deformation tensor, say $\mathcal{B}_{ab}$. Further this tensor can be split into four fundamental and irreducible tensors as 
	\begin{equation}
		\mathcal{B}_{ab}=\nabla_{b}v_{a}=\dfrac{\Theta}{n-1}~ \eta_{ab}+\sigma_{ab}\sigma^{ab}+\omega_{ab}\omega^{ab}-A_{a}v_{b}\label{eq1},
	\end{equation} $n$ is the dimension of the space-time. Now we discuss the geometrical and physical meaning of the introduced tensors in eq.(\ref{eq1}).
\begin{itemize}
	\item \textbf{$\Theta$} is known as the expansion scalar. It is the trace part of $	\mathcal{B}_{ab}$ i.e, $\Theta=\mathcal{B}^{a}_{a}=\nabla_{a}v^{a}$. It describes the average separation between the geodesic worldlines of the $v_{a}$--congruence, precisely the average expansion/contraction of the associated observers.
	\item $\sigma_{ab}$ is known as the shear tensor defined by $\sigma_{ab}=\dfrac{1}{2}\left(\nabla_{b}v_{a}+\nabla_{a}v_{b}\right)-\dfrac{\Theta}{n-1}\eta_{ab}$. It is symmetric traceless part of $	\mathcal{B}_{ab}$ i.e, $\sigma_{ab}=\sigma_{ba}$ and $\sigma^{a}_{a}=0$. It measures the kinematic anisotropies.
	\item $\omega_{ab}$ is the anti-symmetric part of $	\mathcal{B}_{ab}$ (i.e, $\omega_{ab}=-\omega_{ba}$) and is defined by $\omega_{ab}=\dfrac{1}{2}\left(\nabla_{b}v_{a}-\nabla_{a}v_{b}\right)$. It is called the rotation/ vorticity tensor as it measures the kinematic rotation or monitors the rotational behavior of the $v_{a}$--vector field.
	\item $\eta_{ab}=g_{ab}\pm v_{a}v_{b}$ is called the induced metric/projection tensor that operates on the $(n-1)$ dimensional hyper-surface. $\eta_{ab}$ satisfies the orthogonality condition i.e, $v^{b}\eta_{ab}=0$. In expression for $\eta_{ab}$, `+' sign is for time-like curves ($v_{a}v^{a}=-1$) and `-' sign is for null curves ($v_{a}v^{a}=0$).
	\item $A_{a}$ is the 4-acceleration vector field defined by $A_{a}=v^{b}\nabla_{b}v_{a}$. This field guarantees the presence of non gravitational forces. Therefore $A_{a}=0$ for geodesic worldlines.
\end{itemize}
Thus the above discussion shows that the expansion, rotation and shear are purely geometric characteristic of the cross sectional area enclosing a bundle of curves orthogonal to the flow lines. The shape of this area changes as one moves from one point to another along the flow but it still includes the same set of curves in the bundle. The thing that may change during the flow is that the bundle may be isotropically smaller or larger, sheared or twisted. This situation can be made analogous to the elastic deformations or fluid flow. \\
Raychaudhuri equation \cite{Burger:2018hpz}, \cite{Bhattacharyya:2021djv} is nothing but the proper time ($\tau$) evolution of the expansion scalar ($\Theta$)  as
\begin{equation}
	\dfrac{d\Theta}{d\tau}=-\dfrac{\Theta^{2}}{n-1}-\sigma_{ab}\sigma^{ab}+\omega_{ab}\omega^{ab}+\nabla_{b}A^{b}-R_{ab}v^{a}v^{b}\label{eq2**}
\end{equation}
Raychaudhuri equation for null geodesic congruence is given by
\begin{equation}
		\dfrac{d\Theta}{d\lambda}=-\dfrac{\Theta^{2}}{n-2}-\sigma_{ab}\sigma^{ab}+\omega_{ab}\omega^{ab}+\nabla_{b}A^{b}-R_{ab}k^{a}k^{b}\label{eq2*}
	\end{equation} where $k^{a}$ is a null vector and $\lambda$ is an affine parameter.
Ricci tensor $R_{ab}$ is a $(0,2)$ tensor that carries the effect of local gravitational field as otherwise RE is a purely geometric identity and has nothing to do with gravity. The term $-R_{ab}v^{a}v^{b}$ encapsulates the contribution of space-time geometry and is independent of the derivatives of the vector field. Thus in comparison to the other terms present in eq. (\ref{eq2**}), this particular term possesses more general implications. Also, it has a geometrical interpretation as a mean curvature in the direction of the $v_{a}$--congruence.
	\subsection{An outline of the geometric derivation of the Raychaudhuri equation}
	Raychaudhuri addressed the issue of singularity in the early 1950's. At that time he was working on the features of electronic energy bands in metals. The generic features of GR and nature of gravitational singularities fascinated him. Later, motivated by cosmology he pointed out for the first time in his seminal paper that singularity is nothing more than an artifact of the symmetries of the matter distribution. Actually he wanted to see the effect of spin (non-zero vorticity), anisotropy (shear) or cosmological constant can succeed in avoiding the initial singularity. We first review the outline of the derivation of RE from his seminal 1955 paper. The motivation behind the original derivation was entirely devoted to cosmology. Raychaudhuri did not assume homogeneity or isotropy but a time dependent geometry which characterizes a universe i.e, he proposed a time dependent model of the Universe without assuming cosmological principle.The entire derivation was carried out in the synchronous/comoving frame--the frame in which the observer is at rest in the fluid. The space-time coordinates in the 1955 paper were labeled as $x^{1},x^{2},x^{3},x^{4}$, where $x^{^{4}}$ is $t$, the time coordinate. The quantity $R^{4}_{4}$ was evaluated both by using Einstein's field equations with $\Lambda$ (cosmological constant) and again using the geometric definition of $R^{4}_{4}$ in terms of the metric and its derivatives. In order to do so, he used the geometric definition of $\Theta$, $\sigma$ and $\omega$. Consequently, he equated these two ways of writing $R^{4}_{4}$ to get the evolution of expansion--which is the celebrated RE. Motivated by this 1955 paper, many relativists and cosmologists published innumerable papers following Raychaudhuri's work. For example, Heckmann and Schucking \cite{h and s} in the same year derived a set of equations while dealing with Newtonian cosmology. One of the equations resembled the RE in Newtonian case. RE further showed a relativistic generalization of their work without any scientific issue. Later in 1961, Jordan et.al extensively wrote an article on the relativistic mechanics of continuous media where the derivation of these kinematic quantities appeared for the first time.
	\section{Application of Raychaudhuri equation in Relativity and Cosmology}
	\subsection{Geodesic Focusing Theorem}
	The general form of the RE (\ref{eq2**}) can be reduced to much simpler form if one assumes\\
	1. The congruence of curves to be time-like geodesic. In that case $A^{b}=0$.\\
	2. Congruence of time-like geodesics to be hyper surface orthogonal, which by virtue of Frobenius theorem \cite{Poisson:2009pwt} of differential geometry implies zero rotation i.e, $\omega_{ab}=0$. If one is interested in geodesic focusing then zero vorticity congruence must be taken into consideration to avoid centrifugal forces.\\
	Thus the simplified version is 
	\begin{equation}
		\dfrac{d\Theta}{d\tau}=-\dfrac{\Theta^{2}}{n-1}-2\sigma^{2}-\tilde{R}\label{eq3*}
	\end{equation} where $2\sigma^{2}=\sigma_{ab}\sigma^{ab}$ and $\tilde{R}=R_{ab}v^{a}v^{b}$. 
If the matter satisfies the SEC i.e ,
\begin{equation}
	T_{ab}v^{a}v^{b}+\dfrac{1}{2}T\geq0 ,
\end{equation}
then Einstein's equation 
\begin{equation}\label{eq3}
	R_{ab}-\dfrac{1}{2}Rg_{ab}=T_{ab}
\end{equation}
yields, \begin{equation}\label{eq77*}
	R_{ab}v^{a}v^{b}\geq0.
\end{equation}
Employing the condition (\ref{eq77*}) on (\ref{eq3*}) we get ,
\begin{equation}
	\dfrac{d\Theta}{d\tau}+\dfrac{\Theta^{2}}{n-1}\leq0.
\end{equation}
Integrating the above inequality w.r.t proper time $\tau$ we get,
\begin{equation}
	\dfrac{1}{\Theta(\tau)}\geq\dfrac{1}{\Theta_{0}}+\dfrac{\tau}{n-1}.
\end{equation}
Thus, one can infer that any initially converging hyper-surface orthogonal congruence of time-like geodesics must continue to converge within a finite value of the proper time $\tau\leq-(n-1)\Theta_{0}^{-1}$ which leads to crossing of geodesics and formation of a congruence singularity (may or may not be a curvature singularity). This is called the Focusing Theorem (FT) and the condition (\ref{eq77*}) is the corresponding Convergence Condition (CC). Further, it is to be noted that the Strong Energy Condition (SEC) causes gravitation to be attractive and hence can't cause geodesic deviation, rather it increases the rate of convergence. Thus the FT inevitably proves the generic existence of singularity as a major drawback of Einstein gravity. As clear from the above discussion, FT follows as a consequence of the RE, this is the reason why RE is regarded as the fundamental equation of gravitational attraction. Focusing Theorem in terms of null geodesic can be derived in a similar manner by considering the RE in case of null geodesic congruence.
\subsection{Focusing theorem for non zero curvature}
	Focusing Theorem for the flat model of universe is well known in literature. However if one includes curvature then what happens? Motivated by this question, in this section we try to rewrite the Focusing theorem for closed and open model of the universe, sign of curvature being positive and negative respectively. The first Friedmann equation with non-zero curvature term $K$ is given by
\begin{equation}
	3\left(H^{2}+\dfrac{K}{a^{2}}\right)=\rho.\label{eq11*}
\end{equation}
In 4 dimensional FLRW background the expansion scalar is  $\Theta=3H$ where $H=\dfrac{\dot{a}}{a}$ is the Hubble parameter and $a(t)$ is the cosmic scale factor. Writing (\ref{eq11*}) in terms of $\Theta$ we have
\begin{equation}
	\dfrac{\Theta^{2}}{3}=\rho-\dfrac{3K}{a^{2}}.
\end{equation} Using the RE in FLRW model ($\sigma=0$) considering hyper-surface orthogonal ($\omega=0$) congruence of time-like geodesic ($\nabla_{b}A^{b}=0$)  and $\tilde{R}=\frac{1}{2}(\rho+3p)$ one has
\begin{equation}
	\dfrac{d\Theta}{dt}=\dfrac{3K}{a^{2}}-\dfrac{3}{2}(\rho+p)\label{eq13*}
\end{equation}
Now we discuss the following cases:\\
\textbf{Case-I : $K<0$} \\
If $(\rho+p)\geq0$ then equation (\ref{eq13*}) gives $\dfrac{d\Theta}{dt}<0$. Thus the expansion of the congruence decreases with time. In other words if we consider an open model of universe with matter satisfying the Null Energy Condition (NEC) then congruence may focus either in finite or in infinite time. This may lead to formation of congruence singularity. However if $(\rho+p)<0$ and $|\rho+p|>\dfrac{3K}{a^{2}}$ then $\dfrac{d\Theta}{dt}>0$. Therefore open universe with exotic matter can avoid focusing and hence singularity formation.\\ \\
\textbf{Case II: $K>0$} \\ In this case $\dfrac{K}{a^{2}}>0$ and using equation (\ref{eq13*}) one may find that if  $(\rho+p)>\dfrac{2K}{a^{2}}>0$ then $\dfrac{d\Theta}{dt}<0$. Therefore in a closed model of universe if matter satisfies the NEC with a non zero lower bound then expansion of the congruence decreases with time. However singularity may be avoided even with the assumption of NEC on matter provided $\dfrac{2K}{a^{2}}>(\rho+p)$. Thus a high positive curvature and matter content satisfying NEC is required in a closed universe for the possible avoidance of singularity. But in any case exotic matter always avoids the singularity.
	\subsection{An insight of the Singularity theorems}
	Although AKR pointed out the inherent existence of singularities via his celebrated equation in the original 1955 paper, yet more general results which were based on global techniques in Lorentzian spacetimes appeared as singularity theorems in Penrose's work and Hawking's contribution \cite{Penrose:1964wq},\cite{Hawking:1970zqf},\cite{Hawking:1973uf}. The key ingredient of the singularity theorems is that existence of singularity was established by considering Lorentz signature metrics and causality, existence of trapped surfaces and energy conditions on matter. Most importantly, the precise definition of singularity first appeared in their works. The two concepts---notion of geodesic incompleteness and singularities (not necessarily curvature singularities) were aligned. It is to be noted that FT and RE could be completely benign as these are true irrespective of any singularity actually occurring in the space-time manifold. Thus this gives us the knowledge that singularity would always imply focusing but focusing alone cannot imply a singularity. This was pointed out by Landau \cite{Landau:1975pou}. For more details regarding the singularity theorems, there are some phenomenal works \cite{Wald:1984rg},\cite{Hawking:1973uf},\cite{Joshi:1987wg} in this direction for general readership. 
	\subsection{Raychaudhuri scalar in cosmology}
	$\tilde{R}$ or the Raychaudhuri scalar has a cosmological interpretation regarding the convergence of the congruence of time-like geodesics (also known as focusing of geodesics) and in the avoidance of singularity as follows:\\
	In Einstein gravity or in usual modified gravity the field equations for gravity can be written as
	\begin{equation}
		G_{ab}=\kappa~ T_{ab}^{'}~,
	\end{equation}
	where $T_{ab}^{'}=T_{ab}$ is the usual energy-momentum tensor for the matter field in Einstein gravity while $T_{ab}^{'}=T_{ab}+T_{ab}^{(e)}$ for most of the modified gravity theories with $T_{ab}^{(e)}$ containing the extra geometric/physical terms in the field equations. Thus the Raychaudhuri scalar/ Curvature scalar $\tilde{R}$ takes the following form in terms of the energy-momentum tensor or/and the effective energy-momentum tensor as:
	\begin{equation}
		\tilde{R} =\kappa~(T_{ab}-\frac{1}{2}Tg_{ab})v^{b} v^{b}=\dfrac{1}{2}\left(\rho+3p\right),~Einstein~ Gravity\label{eq6*}
	\end{equation}
	\begin{eqnarray}
		\scriptsize	=\small \kappa~\left[(T_{ab}-\frac{1}{2}Tg_{ab})v^{a} v^{b}+(T^{(e)}_{ab}-\frac{1}{2}T^{(e)}g_{ab})v^{a} v^{b}\right]\nonumber\\
		\scriptsize	=\dfrac{1}{2}\left(\rho+3p\right)+\dfrac{1}{2}\left(\rho^{(e)}+3p^{(e)}\right),~Modified ~Gravity\label{eq7***}
	\end{eqnarray}
	Here we show a three fold interpretation of $\tilde{R}$ which lead us to the same conclusion regarding the convergence and avoidance of singularity.
	\begin{enumerate}
		\item 
		In FLRW space-time the (effective) Einstein field equations are
		\begin{eqnarray}
			\scriptsize 	3H^{2}=\kappa \rho,~~2\dot{H}=-\kappa(\rho+p) \label{eq16*}\\
			\scriptsize	3H^{2}=\kappa(\rho+\rho_{e}),~~2\dot{H}=-\kappa[(\rho+p)+(\rho_{e}+p_{e})]\label{eq7**}
		\end{eqnarray} where equation (\ref{eq16*}) is for Einstein gravity, (\ref{eq7**}) corresponds to Modified gravity and `.' is differentiation w.r.t cosmic time $t$.\\
		So the deceleration parameter $q=-\left(1+\dfrac{\dot{H}}{H^{2}}\right)$ takes the form:
		\begin{eqnarray}
			q=\dfrac{\rho+3p}{2\rho}, ~Einstein~ gravity.\nonumber\\
			~~~~q=\dfrac{(\rho+3p)+(\rho_{e}+3p_{e})}{2(\rho+\rho_{e})},~ Modified ~gravity.\nonumber
		\end{eqnarray}
		Hence 
		\begin{eqnarray}
			\tilde{R}=q\rho~, Einstein~gravity\nonumber
			\\ 
			\tilde{R}=q(\rho+\rho_{e})~, Modified~gravity\nonumber
		\end{eqnarray} One finds that $\tilde{R}=\dfrac{3qH^{2}}{\kappa}$ for both the cases. Now for convergence $\tilde{R}>0$ so one may conclude that convergence will occur during the evolution of the universe if $q>0$. Thus formation of singularity is not possible both in early inflationary era and the present late time accelerated era of evolution, while the matter dominated era of evolution favors the convergence.
		\item From eq (\ref{eq6*}) (i.e in Einstein gravity) if we write $\rho=\rho_{1}+\rho_{2}$ and $p=p_{1}+p_{2}$ then we find that  $\rho_{2}+3p_{2}<0$ and
		$|\rho_{2}+3p_{2}|> \rho_{1}+3p_{1}$  yield $\tilde{R}<0$. In other words $(\rho_{2},~p_{2})$ corresponds to the energy density and thermodynamic pressure of dark energy. Hence dominance of dark energy over normal matter ( having density and pressure ($\rho_{1},~p_{1}$)) does not allow Focusing Theorem to hold. Therefore the era dominated by dark energy namely the inflationary era and the present accelerated era of expansion are against the formation of singularity.
		\item From eq (\ref{eq7***}) (i.e. in modified gravity), to make $\tilde{R}<0$ we need $\rho^{(e)}+3p^{(e)}<0$ and $|\rho^{(e)}+3p^{(e)}|> \rho+3p$. Again it hints that $(\rho^{(e)},p^{(e)})$ corresponds to the density and pressure of dark energy and hence we arrive at the same conclusion as in the former cases.
	\end{enumerate}
	\subsection{Black Hole and Raychaudhuri scalar}
\large	In order to prove the existence of Black-Hole singularity using the Raychaudhuri equation we consider the Schwarzschild metric
	\begin{equation}
		ds^{2}=-\left(1-\dfrac{2GM}{r}\right)dt^{2}+\left(1-\dfrac{2GM}{r}\right)^{-1}dr^{2}+r^{2}d\Omega^{2}\label{eq*1}.
	\end{equation} Schwarzschild metric is a vacuum solution of the Einstein's field equations and in general relativity a vacuum solution is a Lorentzian manifold whose Einstein tensor (and hence the Ricci tensor $R_{ab}$) vanishes identically and thus $\tilde{R}=0$. Therefore the Convergence scalar $R_{c}=\tilde{R}+2\sigma^{2}=2\sigma^{2}$. By the dynamics of classical Schwarzschild metric \cite{Blanchette:2021jcw} and using the definition of anisotropy scalar ($\sigma^{2}$) one has
	\begin{equation}
		R_{c}=2\sigma^{2}=\dfrac{4}{3r^{3}}\dfrac{(3GM-r)^{2}}{(2GM-r)}.\nonumber
	\end{equation} Based on the above expression of $R_{c}$ one has the following findings:\\ \\
	$~~~~~$(i) $R_{c}>0$ only when  $r<2GM$ . So singularity may be possible only when  $r<2GM$. Further as $r\rightarrow0$  a stage will come when $R_{c}$ will predominate over $\omega^{2}$ and $\nabla_{c}A^{c}$ so that $R_{c}-2\omega^{2}-\nabla_{c}A^{c}\geq0$ (Convergence Condition) and this hints that convergence and hence formation of singularity may occur.\\ 
	
	(ii) For $r>2GM$, $R_{c}<0$. So singularity is not possible in the region $r>2GM$. Therefore $r=2GM$ acts a boundary to distinguish the regions with and without singularity. Here $r=2GM$ is nothing but the Event-Horizon.\\
	
	(iii) $r=0$ and $r=2GM$ are the points where $R_{c}$ diverges/ blows off. These are therefore identified as the points of singularity. However singularity at $r=2GM$ is a co-ordinate singularity that arises due to bad choice of coordinates and can be removed in Eddington-Finkelstein co-ordinates. On the other hand $r=0$ is the physical singularity or the Black-Hole singularity whose existence is inevitable from the point of view of the RE.\\ \\
	Therefore RE tells us that even if there is no matter (i.e vacuum) yet the space-time may
	develop singularity due to presence of anisotropy and Black-Hole singularity is an example of
	such a case.
	\subsection{Wormhole and Raychaudhuri Equation}
	Under very general conditions, a traversable wormhole violates the average Null Energy Condition (NEC) in the region of the throat. This can be shown using the Raychaudhuri equation (RE), together with the fact that a wormhole throat by definition defocuses light rays. We know that Focusing theorem is the most vital consequence of the RE which needs $\tilde{R}\geq0$ (CC) for geodesic focusing. If a space-time has a singularity, then a bundle of geodesic will tend to focus at the singularity. Thus a violation of the CC may possibly avoid singularity. The condition that the wormhole be traversable, in particular means that there are no event horizons or curvature singularities. Now to construct a traversable wormhole we consider the Morris-Thorne line element given by \cite{Morris:1988cz},\cite{Lemos:2003jb},\cite{DeFalco:2023twb}
	\begin{equation}
		ds^{2}=-e^{2\phi(r)} dt^{2}+\left(1-\dfrac{b(r)}{r}\right)^{-1}dr^{2}+r^{2}d\Omega_2^{2}\label{eq19*}
	\end{equation} where
\begin{equation}
	d\Omega_2^{2}=d\theta^{2}+\sin^{2}\theta d\phi^{2}
\end{equation} $\phi(r)$ is the redshift function and $b(r)$ is the shape function. For the line element (\ref{eq19*}), the Einstein tensor components are 
\begin{eqnarray}
	G_{00}=\dfrac{b'(r)}{r^{2}}\\
	G_{11}=-\dfrac{b}{r^{3}}+2\left(1-\dfrac{b(r)}{r}\right)\dfrac{\phi'}{r}\\
	G_{22}=G_{33}=\left(1-\dfrac{b(r)}{r}\right)\left(\phi''+\phi'^{2}+\left(\dfrac{-rb'+2r-b}{2r(r-b)}\right)\phi'-\dfrac{(rb'-b)}{2r^{2}(r-b)}\right)
\end{eqnarray}
The RE for hyper-surface orthogonal null geodesics is given by 
\begin{equation}
	\dfrac{d\Theta}{d\lambda}=-\dfrac{\Theta^{2}}{2}-2\sigma^{2}+R_{\mu\nu}k^{\mu}k^{\nu}
\end{equation} where $k^{\mu}$ is the null vector so that $k_{\mu}k^{\mu}=0$. In this context the CC becomes $R_{\mu\nu}k^{\mu}k^{\nu}\geq0$. Now from the Einstein's field equation
\begin{equation}
	G_{\mu\nu}=R_{\mu\nu}-\dfrac{1}{2}R g_{\mu\nu}
\end{equation} one has $\tilde{R}=R_{\mu\nu}k^{\mu}k^{\nu}=G_{\mu\nu}k^{\mu}k^{\nu}$ using the property of null vector $k^{\mu}$. For infalling observer we can take the null vector as $(\sqrt{-g^{00}},\pm \sqrt{g^{11}},0,0)$
i.e, $k^{0}=e^{\phi}$, $k^{1}=\pm \dfrac{1}{\left(1-\dfrac{b(r)}{r}\right)^{\frac{1}{2}}}$,  $k^{2}=k^{3}=0$. The Raychaudhuri scalar turns out to be
\begin{equation}
	\tilde{R}=G_{\mu\nu}k^{\mu}k\nu=\dfrac{b'}{r^{2}}e^{\phi}-\dfrac{b(r)}{r^{3}(1-\frac{b(r)}{r})}+\dfrac{2}{r}\phi'
\end{equation}[\textbf{The consequences with time-like geodesic has been discussed in the Appendix}]. Now to have traversable wormhole we need the following:
\begin{itemize}
	\item Wormhole is a bridge between two asymptotically flat regions connected by a throat. The throat radius is defined by a global minimum $r=r_{0}$, so the radial coordinate varies as $r\geq r_{0}$.
	\item The redshift function $\phi(r)$ must be finite everywhere in order to avoid the presence of horizons and singularity. So, $e^{\phi(r)}>0$ and finite for $r>r_{0}$. In this context, the ultrastatic wormhole is a particular point of interest which defines the zero-tidal force wormhole. In case of ultrastatic wormhole, $\phi(r)=0$ so that $e^{2\phi(r)}=1$. Thus 
	\begin{equation}
	\tilde{R}=\dfrac{b'}{r^{2}}-\dfrac{b(r)}{r^{2}(r-b)}.
	\end{equation}
	\item Flairing out condition: $\left(\dfrac{-rb'(r)+b(r)}{b^{2}(r)}>0\right)$ must hold near the throat.
	\item At $r=r_{0}$, $b(r)=r$, $b'(r)=1$ and $r>r_{0}$ implies $b(r)<r$ for all $r>r_{0}$.
	\item The asymptotic flatness implies that $\phi(r)\rightarrow0$ and $\dfrac{b(r)}{r}\rightarrow0$ as $r\rightarrow \infty$
\end{itemize}
Thus, writing $\tilde{R}=\dfrac{(r-b)b'-b}{r^{2}(r-b)}=-\dfrac{(-rb'+b)+bb'}{r^{2}(r-b)}$. We have seen that for $r>r_{0}$, $(r-b)>0$ thus denominator of $\tilde{R}$ is positive. To satisfy the flair out conditions we need $(rb'-b)<0$. Also we have $bb'>0$. This is because, $b(r)=r>0$ for all $r\geq r_{0}$ and $b'(r)\leq1$ for all $r\geq r_{0}$. This shows that the numerator of $\tilde{R}$ is negative. Hence CC is violated near the throat and there is defocusing of null geodesics. This is illustrated in the following example where we consider an incoming light travelling along geodesics. It then crosses a wormhole and again expands on the other side of the wormhole. Since the neck of the wormhole is of finite length, thus Focusing theorem does not hold there and hence formation of singularity at least within the vicinity of the neck is forbidden. According to the optical Raychaudhuri's theorem this phenomena requires a violation of the NEC. To speak lucidly, the existence of a traversable wormhole is plausible only if the geodesics entering the wormhole on one side (and thus converging as they approach the throat) will emerge on the other side diverging away from each other. By Raychaudhuri's equation this is possible if certain energy conditions are violated. Thus the RE hints why exotic matter or configurations which violate energy conditions are needed for a traversable wormhole to exist. The above statements can be illustrated by considering an example of a wormhole described by the metric \cite{Visser:1995cc}
	\begin{equation}
		ds^{2}=-dt^{2}+dr^{2}+(b^{2}+r^{2})(d\theta^{2}+\sin^{2}\theta d\phi^{2}),~-\infty<t<+\infty,~-\infty<r<+\infty
	\end{equation}
The space-time associated with this metric is asymptotically flat and hosts a stable wormhole of radius $b$ at $r=0$, the origin. The metric does not possess a singularity nor there is an existence of event horizon. Now we see the implications of RE or more precisely the Raychaudhuri scalar for this particular type of wormhole. The expression for $\tilde{R}$ turns out to be
\begin{equation}
\tilde{R}=R_{ab}v^{a}v^{b}=R_{rr}(v^{r})^{2}
\end{equation} as the only non zero component of Ricci tensor is $R_{rr}=-\dfrac{2b^{2}}{(b^{2}+r^{2})^{2}}$. Thus $\tilde{R}<0$ and CC is violated. This proves that there is no singularity. This is because, if there were a singularity focusing would have occured there. Also $\tilde{R}<0$ stands for violation of energy conditions. Thus the theoretical description of a traversable wormhole from the point of view of RE matches with the implication of RE in the above wormhole. This shows that the above wormhole is a traversable one and in literature, it is named as Ellis Wormhole. Thus, RE identifies the existence of a traversable wormhole. One can find the Raychaudhuri scalar $\tilde{R}$ and expansion scalar $\Theta$ for a general wormhole metric and see the consequences of RE in terms of average energy conditions. Wormholes are a case where a series of singularity theorems break due to violation of certain energy conditions. Thus not only black hole but also RE is equally important in the study of traversable wormholes. 
	\subsection{Raychaudhuri equation in Modified Gravity theories}
	\subsubsection{$f(R)$ gravity in inhomogeneous FLRW model: Raychaudhuri equation}
\large	The usual Einstein-Hilbert action In $f(R)$ gravity theory \cite{Nojiri:2010wj},\cite{Bamba:2012cp},\cite{DeFelice:2010aj}
 is generalized as
	\begin{equation}
		\mathcal{A}=\dfrac{1}{2\kappa}\int\sqrt{-g}f(R)\mathrm{d}^4x+\int\mathrm{d}^4x\sqrt{-g}\mathcal{L}_{m}(g_{ab},\Sigma).\label{eq21***}
	\end{equation}
	Here $\mathcal{L}_{m}$ is the Lagrangian of the matter field with $\Sigma$ denoting the coupling between matter source and geometry $(g_{ab})$, $f(R)$ is a continuous function of the Ricci scalar $R$ and $\kappa=8{\pi}G=c=1$ is the usual gravitational coupling constant. Therefore, by varying the above action (\ref{eq21***}) with respect to $g_{ab}$,  the field equations for $f(R)$-gravity can be written in compact form as 
	\begin{equation}\label{eq24*}
		F(R)R_{ab}-\frac{1}{2}f(R)g_{ab}-\left(\nabla_\mu\nabla_\nu-g_{ab}\Box\right)F(R)=\kappa T_{ab},
	\end{equation}
	with $F(R)=\dfrac{\mathrm{d}f(R)}{\mathrm{d}R}=f'(R)$  and \begin{equation}T_{ab}=\dfrac{2}{\sqrt{-g}}\dfrac{\partial\left(\sqrt{-g}\mathcal{L}_{m}\right)}{\partial{g}^{ab}},
	\end{equation} denotes the stress energy tensor of the matter field. The D'Alembertian Operator $\Box$ can be written as 
	$\Box=g^{ab}\nabla_{\mu}\nabla_{\nu}$.
	The trace of the field equation (\ref{eq24*}) gives,
	\begin{equation}\label{eq26*}
		3\Box F(R)+RF(R)-2f(R)=\kappa T.
	\end{equation}
	
	Now, combining the field equation (\ref{eq24*}) with (\ref{eq26*}) gives the modified Einstein field equations (after some algebraic manipulation) as \cite{Bhattacharya:2015oma}
	\begin{equation}
		G_{ab}=\tilde{T}_{ab}+\dfrac{1}{F}(\nabla_{\mu}\nabla_{\nu}F-g_{ab}N),
	\end{equation}
	where
	\begin{equation}
		\tilde{T}_{ab}=\dfrac{1}{F}T_{ab},  ~~ N(t,r)=\dfrac{1}{4}(RF+\Box F+T).
	\end{equation}  Now from the field equations (\ref{eq24*}) and their trace equation (\ref{eq26*}) one gets for a unit time-like vector $v^{\mu}$,
	\begin{equation}\label{eq7****}
		\tilde{R}=	R_{ab}v^{a}v^{b}=\dfrac{1}{F(R)}\left[-\dfrac{1}{2}\Box F(R)+\dfrac{1}{2}({f(R)-RF(R)})+\kappa\left(T_{ab}v^{a}v^{b}+\dfrac{1}{2}T\right)+v^{a}v^{b}\nabla_{a}\nabla_{b}F(R)\right],
	\end{equation}
	
	So for CC of a congruence of time-like curves having $v^{a}$ as the unit tangent vector field, the r.h.s of the above equation must be positive semi-definite. In the background of inhomogeneous FLRW space-time geometry having line-element \cite{Bhattacharya:2016env}
	\begin{equation}\label{eq8}
		\mathrm{d}s^2=-\mathrm{d}t^2+a^2(t)\left[\frac{\mathrm{d}r^2}{1-\frac{b(r)}{r}}+r^2\mathrm{d}\Omega_2^2\right],
	\end{equation}
	where $\mathrm{d}\Omega_2^2=d\theta^{2}+\sin^{2}\theta d\Phi^{2}$ is the metric on 2-sphere ($\theta$ being the polar angle),  $a(t)$ is the scale factor, $b(r)$ is an arbitrary function of $r$, the scalar curvature has the form
	\begin{equation}
		R=6(\dot{H}+2H^{2})+2\dfrac{b'}{a^{2}r^{2}}.
	\end{equation}
	
\large	The field equations for $f(R)$-gravity has the explicit form
	\begin{eqnarray}
		3H^2+\frac{b'(r)}{a^2r^2}&=&\frac{\rho(r,t)}{F(R)}+\frac{\rho_e(r,t)}{F(R)}\label{eq28*}\\
		-\left(2\dot{H}+3H^2\right)-\frac{b(r)}{a^2r^3}&=&\frac{p_r(r,t)}{F(R)}+\frac{p_{re}(r,t)}{F(R)}\label{eq33}\\
		-\left(2\dot{H}+3H^2\right)-\frac{(b-rb')}{2a^2r^3}&=&\frac{p_t(r,t)}{F(R)}+\frac{p_{te}(r,t)}{F(R)}\label{eq28**}
	\end{eqnarray}
	where $H=\dfrac{\dot{a(t)}}{a(t)}$ is the usual Hubble parameter. Here the matter is in the form of cosmic anisotropic fluid with $\rho=\rho(r,t)$, $p_{r}=p_{r}(r,t)$, $p_{t}=p_{t}(r,t)$ as the energy density, radial and transverse pressures respectively. Also the expression for the hypothetical matter components are 
	\begin{equation}
		\rho_e=N+\ddot{F} , ~ p_{re}=-N-H\dot{F}+\frac{(r-b)}{a^2r}F'-\frac{(b-rb')}{2a^2r^2}F', ~p_{te}=-N-H\dot{F}+\frac{(r-b)}{a^2r^2}F',
	\end{equation} 
	
	Now the conservation relations for the anisotropic fluid can be written as :
	\begin{equation}\label{eq32****}
		\frac{\partial\rho}{\partial t}+(3\rho+p_r+2p_t)H=0,\mbox{~ and ~}
		\frac{\partial p_r}{\partial r}=\frac{2}{r}(p_t-p_r)
	\end{equation}
	
\large	Although from cosmological principle the universe is on the large scale homogeneous and isotropic and almost all models in cosmology has this property due to the elegance and simplicity of these models, still the universe is not fundamentally homogeneous on the scale of galaxies clusters and superclusters -- there is clumping of matter. Further, at the very early phase of the universe, it is very likely to have a state of much disorder (for example in emergent era/inflationary epoch). Moreover, it is possible to have apparent acceleration of the universe due to the back reaction on the metric of the local inhomogeneities \cite{Pascual-Sanchez:1999xpt,Rasanen:2006kp}. Thus it is reasonable to consider the inhomogeneous model as an alternative to dark energy. For the present $f(R)$ gravity model it is reasonable to choose 
	\begin{equation}\label{eq15**}
		b(r)=b_{0}\left(\dfrac{r}{r_{0}}\right)^{3}+d_{0}=\mu_{0}r^{3}+d_0 
	\end{equation}so that
	\begin{equation}
		R=6(\dot{H}+2H^{2})+\frac{6\mu_{0}}{a^{2}}\label{eq34**}
	\end{equation}
\large	is a function of `$t$' alone. This choice of $b(r)$ includes two parameters, namely $\mu_{0}$ and $d_{0}(\neq0)$, where $d_{0}$ is identified as the inhomogeneity parameter. The motivation behind choosing a suppressed 3rd order polynomial (not any general degree polynomial or other analytic function) lies in the fact that this particular choice of $b(r)$ with $d_{0}=0$ reduces the present model to FLRW and hence successfully helps us to study an inhomogeneous model as an alternative to dark energy. Moreover with this choice of $b(r)$, the scalar curvature $R$ given by (\ref{eq34**}) turns out to be homogeneous which leads to a lot of simplification in mathematical calculations. The above modified Friedmann equations (\ref{eq28*})-(\ref{eq28**}) with $b(r)$ from equation (\ref{eq15**}) has a possible solution for the two matter components as \small
	\begin{eqnarray}
		\rho=3H^2g(t) ~,~ \rho_e=\frac{3\mu_0g(t)}{a^{2}}&,& p_{re}=p_{te}=-\frac{\mu_0}{a^2}+H\dot{g(t)} , \nonumber\\ p_r=\psi(t)\left[-\left(2\dot{H}+3H^2\right)-\frac{d_0}{a^2r^3}\right]-H\dot{g}(t) &,&		p_t=g(t)\left[-\left(2\dot{H}+3H^2\right)+\frac{d_0}{2a^2r^3}\right]-H\dot{g}(t)\label{eq35*}
	\end{eqnarray}
	
\large	The above choice shows that the usual matter component is inhomogeneous and anisotropic in nature while the hypothetical curvature fluid is both homogeneous and isotropic in nature with $\omega_{e}=\dfrac{1}{3}\left(1+\dfrac{Ha^{2}\dot{g(t)}}{\mu_{0}}\right)$ as the expression for state parameter. Further, due to the inhomogeneity (i.e $d_{0}\neq0$) the equation of state parameters for the normal fluids i.e $\omega_{r}=\dfrac{p_{r}}{\rho}$ and $\omega_t=\dfrac{p_{t}}{\rho}$ are related linearly as 
	\begin{equation}\label{eq40}
		\omega_t-\omega_r=\frac{d_0}{2a^2H^2r^3}
	\end{equation}
	
	It may be noted that both  the conservation equations given by (\ref{eq32****}) will be satisfied identically for the choice 
	\begin{equation}\label{eq19}
		\omega_t=\frac{d_0}{6a^2H^2r^3}~ , ~ \omega_r=-\frac{d_0}{3H^2a^2r^3}
	\end{equation} 
	Moreover, this choice of the state parameters results a differential equation in $\psi(t)=F(R)$ as 
	\begin{equation}
		\dfrac{\dot\psi}{\psi}+2\dfrac{\dot{H}}{H}+3H=0,
	\end{equation}
	which has the solution 
	\begin{equation}
		\psi=\frac{\psi_{0}}{(a^{3}H^{2})}
	\end{equation}
	This solution shows that $f(R)$ will be in the power-law form of $R$ if the power-law form of expansion of the universe is assumed. Subsequently, in the following subsections we have used the power-law form of $f(R)$ particularly in the study of the CC. Also the power-law form of $f(R)$ over Einstein gravity is suitable for inflationary scenario. However, from the point of view of hypothetical curvature fluid the different components of the fluid (given by equation (\ref{eq35*})) results a differential equation for $\psi$ as
	\begin{equation}
		\ddot\psi-2H\dot\psi-\dfrac{2\mu_{0}}{a^{2}r^{3}}\psi=0,
	\end{equation}
	having solution of the form (with $\mu_{0}=0$)
	\begin{equation}
		\psi(t)=\dfrac{\psi_{0}}{2}\int\dfrac{d(a^{2})}{H}
	\end{equation}
	Here also $f(R)$ is of the form $R^{-(n+\frac{1}{2})}$ for power law expansion of the universe.
	Finally for the choice (\ref{eq15**}) of the given inhomogeneous fluid after a little bit of algebra with the field equations one obtains the RE as
	\begin{equation}
		\frac{\ddot{a}}{a}=-\frac{1}{2\psi(t)}\left[\frac{\rho}{3}+H\dot{\psi}\right]
	\end{equation}
	Thus the RE is a homogeneous equation although the spacetime geometry is inhomogeneous in nature. Further, the RE does not depend on the equation of state parameters, it depends only on the energy density of the physical fluid. 
	\subsubsection{Convergence Condition in $f(R)$ modified gravity theory in inhomogeneous background}
\large	Convergence condition ($R_{ab}v^{a}v^{b}\ge0$) for a congruence of time-like curves in the present inhomogeneous space-time has been discussed in details in the paper \cite{Chakraborty:2023ork}. If $v^{a}=(1,0,0,0)$ denotes the unit time-like vector field along the congruence then from (\ref{eq7****}) ,  for CC we must have
	\begin{equation}
		R_{ab}v^{a}v^{b}=\dfrac{1}{F(R)}\left[T_{1}+T_{2}+T_{3}\right] \ge0
	\end{equation}
	where 
	\begin{eqnarray}
		T_{1}&=&\kappa\left(T_{ab} v^{a}v^{b}+\dfrac{1}{2}T\right)\\
		T_{2}&=&-\dfrac{3}{2}\Box{F(R)}=\dfrac{3}{2}v^{a}v^{b}\nabla_{a}\nabla_{b}F(R)\\
		T_{3}&=&\dfrac{1}{2}(f(R)-RF(R))
	\end{eqnarray}
	Note that if the matter satisfies SEC then $T_{1}\ge0$ for all time. The time variation of $T_{1}$, $T_{2}$, $T_{3}$ and $R_{ab}v^{a}v^{b}$ has been shown graphically in the paper \cite{Chakraborty:2023ork} considering the cosmic fluid to be perfect fluid with barotropic equation of state $p=\omega\rho$. The figure shows that the CC is not universally satisfied, it depends on the choice of the parameters involved. Thus it is possible to avoid the singularity in the present model. Hence, existence of singularity is not a generic one but depends on the gravity theory and the space-time background under consideration. Though the present model is inhomogeneous but still the RE so constructed turns out to be a homogeneous differential equation. For congruence of time-like geodesic , convergence  conditions have been analyzed graphically and one may conclude that singularity may be avoided for specific choices for the parameters involved.
	\subsubsection{$f(T)$ gravity in homogenenous FLRW model: Raychaudhuri equation}
	\large	Let us begin with the action for $f(T)$ gravity \cite{Chen:2010va},\cite{Cai:2015emx}
	\begin{equation}\label{eq1*}
		\mathcal{A}_{m}=\dfrac{1}{2}\int\mathrm{d}^{4}x {~}e\left[T+f(T)+\mathcal{L}_{m}\right]
	\end{equation}
	where $T$ is the torsion scalar , $f(T)$ is an arbitrary differentiable function of the torsion $T$, $e=\sqrt{-g}= det (e^{\mathcal{A}_{m}}_{a})$, and $\mathcal{L}_{m}$ corresponds to the matter Lagrangian, $\kappa=8\pi G=1$.\\
	The above torsion scalar $T$ is defined as 
	\begin{equation}\label{eq2}
		T=S_{\sigma}^{ab}{~} T_{ab}^{\sigma}
	\end{equation}
	where  $T^{\sigma}_{ab}$, the torsion tensor is defined as 
	\begin{equation}\label{eq3}
		T^{\sigma}_{ab} = \Gamma^{\sigma}_{ba}-\Gamma^{\sigma}_{ab}=\\e^{\sigma}_{A}{~}(\partial_{a}e^{A}_{b}-\partial_{b}e^{A}_{a})
	\end{equation}
	The Weitzenbock connection $\Gamma^{\sigma}_{ab}$ is defined as 
	\begin{equation}
		\Gamma^{\sigma}_{ab}= e^{\sigma}_{A} {~}\partial_{b}e^{A}_{a} ,
	\end{equation}
	and the super-potential, $S^{ab}_{\sigma}$ is defined as 
	\begin{equation}\label{eq5}
		S^{ab}_{\sigma}=\dfrac{1}{2}\left(K^{ab}_{{~~}\sigma}+\delta^{a}_{\sigma}{~} T^{\alpha b}_{{~~}\alpha} - \delta^{b}_{\sigma}{~}T^{\alpha a}_{{~~}\alpha}\right),
	\end{equation}
	where the contortion tensor $K^{ab}_{{~~}\sigma}$ takes the form
	\begin{equation} \label{eq6}
		K^{ab}_{{~~}\sigma}=-\dfrac{1}{2}\left(T^{ab}_{{~~}\sigma}-T^{ba}_{{~~}\sigma}-T^{ab}_{\sigma}\right).
	\end{equation}
	Geometrically, the orthogonal tetrad components $e_{\mathcal{A}_{m}}(x^{a})$ (considered as dynamical variables), form an orthonormal basis for the tangent space at each point $x^{a}$ of the manifold i.e 
	\begin{equation}\label{eq55}
		e_{i}{\cdot}e_{j}=\eta_{ij} {~~}, {~~}\eta_{ij}= diag{~}(+1,-1,-1,-1)
	\end{equation}
	In a coordinate basis, we may write\\ \begin{equation}
		e_{i}=e^{a}_{i}{~}\partial_{a}
	\end{equation} where $e^{a}_{i}$ are the components of $e_{i}$ , with  $a,{~}i=0,1,2,3$ .
	The metric tensor is obtained from the dual vierbein as 
	\begin{equation}
		g_{ab}(x)=\eta_{ij}{~}e^{i}_{a}(x){~}e^{j}_{b}(x)
	\end{equation}
	The present work deals with $f(T)$ gravity in the framework of homogeneous and isotropic FLRW space-time having line element 
	\begin{equation}
		ds^{2}=-dt^{2}+a^{2}(t)\left[\dfrac{dr^{2}}{1-kr^{2}}+r^{2}(d\theta^{2}+\sin^{2}\theta d\phi^{2})\right]
	\end{equation}
	where $a(t)$ is the scale factor, $H=\dfrac{\dot{a}}{a}$ is the Hubble parameter, '.' denotes differentiation w.r.t cosmic time $t$. \\'$k$', the curvature index dictates the model of the universe. Further it has been assumed that the universe is filled with perfect fluid having barotropic equation of state 
	\begin{equation}\label{eq11}
		p=\omega\rho
	\end{equation}
	where $\omega=\gamma-1{~~} ({~~}0\leq\gamma\leq2{~~})$ being the equation of state parameter.\\ Now using equations (\ref{eq2}), (\ref{eq3}), (\ref{eq5}), (\ref{eq6}) and (\ref{eq55})  we have , 
	\begin{equation}
		T=-6H^{2}
	\end{equation}
	It is to be noted that during cosmic evolution  $T$ is negative. Varying the action (\ref{eq1*}) we get the modified Einstein field equations as
	\begin{equation}\label{eq13}
		\left[	e^{-1}\partial_{a}\left(e{~}e_{A}^{\rho}{~}S_{\rho}^{ab}\right)-e_{A}^{\lambda}{~}T_{a\lambda}^{\rho}{~}S^{ba}_{\rho}\right]\left[1+f_{T}\right]+e_{A}^{\rho}{~}S_{\rho}^{ab}{~}\partial_{a}(T)f_{TT}+\dfrac{1}{4}{~}e_{A}^{b}\left[T+f(T)\right]=4\pi G{~}e_{A}^{\rho}{~}T^{b}_{\rho}
	\end{equation}
	where $f_{T}=\dfrac{df}{dT}$ , $f_{TT}=\dfrac{d^{2}f}{dT^{2}}$ ,  and $T^{b}_{\rho}$ is the energy momentum tensor of the total matter -- baryonic matter and dark energy. Thus for FLRW model, the modified Friedmann equations can be written as :
	\begin{equation}\label{eq62}
		H^{2}=\dfrac{1}{2f_{T}+1}\left[\dfrac{\rho}{3}-\dfrac{f}{6}\right]\end{equation}\begin{equation}\label{eq63}
		2\dot{H}=\dfrac{-(p+\rho)}{1+f_{T}+2Tf_{TT}}
	\end{equation}
	where $p$ and $\rho$ are the thermodynamic pressure and density of the matter fluid having conservation equation
	\begin{equation}\label{eq16}
		\dot{\rho}+3H(p+\rho)=0.
	\end{equation}
	Based on our assumption (\ref{eq11}) , the solution of the differential equation (\ref{eq16}) can be written in the form 
	\begin{equation}\label{eq31}
		\rho=\rho_{0}{~}a^{-3\gamma}.
	\end{equation}
	Finally, using the expression for $H$ and equations (\ref{eq62}), (\ref{eq63}) after some algebraic manipulation one gets the Raychaudhuri equation in $f(T)$ gravity as
	\begin{equation}
		\dfrac{\ddot{a}}{a}=\rho_{0}a^{-3\gamma}\left[\dfrac{1}{3(2f_{T}+1)}-\dfrac{\gamma}{2(1+f_{T}+2Tf_{TT})}\right]-\dfrac{f(T)}{6(2f_{T}+1)}.
	\end{equation}
	Therefore, the Raychaudhuri equation is homogeneous and depends on $\gamma$ which is related to the equation of state parameter $\omega$ by the relation $\gamma=\omega+1$ and choice of the torsion function $f(T)$. This hints that the CC essentially depends on the function $f(T)$.
	\subsubsection{Convergence Condition in $f(T)$ modified gravity}
	The field equations for $f(T)$ gravity given by equations (\ref{eq62}) and ( \ref{eq63}) may be expressed as :
	\begin{eqnarray}
		3H^{2}=(\rho+\rho^{(e)})\\
		2\dot{H}=-(p+\rho)-(p^{(e)}+\rho^{(e)})
	\end{eqnarray}where,
	\begin{eqnarray}\label{eq69}
		\rho^{(e)}=-\left(\dfrac{f(T)}{2}+6H^{2}f_{T}\right)
		\\ \label{eq70}p^{(e)}=2\left(\dot{H}+3H^{2}\right)f_{T}+4Tf_{TT}\dot{H}+\dfrac{f(T)}{2}
	\end{eqnarray} are the energy density and thermodynamic pressure of the effective fluid. The Raychaudhuri scalar $\tilde{R}$=$R_{ab}v^{a}v^{b}$ in this modified gravity turns out to be ,
	\begin{equation}
		R_{ab}v^{a}v^{b}=\left(J_{ab}v^{a}v^{b}+\dfrac{1}{2}J\right)+\left(J_{ab}^{(e)}v^{a}v^{b}+\dfrac{1}{2}J^{(e)}\right),
	\end{equation}
	$J$ being the trace of $J_{\mu\nu}$ i.e $J=g^{ab}J_{ab}$.
	Energy momentum tensor for perfect fluid having unit time-like vector $v^{a}$ ( so that $v^{a}v_{a}=-1$) is given by 
	\begin{equation}
		J_{ab}=(p+\rho)v_{a}v_{b}+pg_{ab}
	\end{equation}
	Thus, the expression for the Raychaudhuri scalar ($\tilde{R}$) is
	\begin{equation}
		\tilde{R}=	R_{ab}v^{a}v^{b}=\dfrac{1}{2}\left(\rho+3p\right)+\dfrac{1}{2}\left(\rho^{(e)}+3p^{(e)}\right)
	\end{equation}
	Using the equations (\ref{eq62}), (\ref{eq63}), (\ref{eq31}), (\ref{eq69}) and (\ref{eq70}), the explicit expression for $\tilde{R}$ in terms of $f(T)$, $f_{T}$ and $f_{TT}$ can be written as
	\begin{equation}
		\tilde{R}=	R_{ab}v^{a}v^{b}=\dfrac{3\gamma\rho_{0}a^{-3\gamma}}{2(1+f_{T}+2Tf_{TT})}+\dfrac{\left(-\rho_{0}a^{-3\gamma}+\frac{f(T)}{2}\right)}{\left(1+2f_{T}\right)}.
	\end{equation}
	Therefore, the expression for the Raychaudhuri scalar ($\tilde{R}$) shows that the CC essentially depends on the choice of $f(T)$. The paper \cite{Chakraborty:2023yyz} illustrates graphically the convergence condition with power law choice for the cosmic scale factor as $a(t)=a_{0}t^{m}$ and finds that for the model $f(T)=\alpha(-T)^{n}$, the CC~ i.e $\tilde{R}\geq0$ holds for  negative $m$ and $n$ while singularity may be avoided for positive exponents. It may be noted that in all cases SEC ($\rho+3p=(3\gamma-2)\rho_{0}t^{-3m\gamma}\geq0$) holds good.
	For the model $f(T)=\alpha T+ \dfrac{\beta}{T}$, $\tilde{R}$ is either positive or indefinite in sign whenever SEC holds good, while negative $\rho_{0}$ or violation of SEC ($(3\gamma-2)\rho_{0}t^{-3m\gamma}<0$) yields $\tilde{R}<0$. Hence singularity may be avoided for any negative choice of $\rho_{0}$ irrespective of the exponent $m$.
	Therefore unlike GR, it is possible to obviate singularity in Model I with suitable choice of the model parameters even with the assumption of SEC. However avoidance of singularity in Model II demands the violation of SEC or hints the ghost nature of matter.
	\subsubsection{Scalar tensor theory}
	In the work \cite{Choudhury:2021zij}, focusing behavior of geodesics and possible avoidance of singularity have been attempted via RE in scalar tensor gravity theory. Two theories namely- Brans-Dicke and Bekenstein's scalar field theory have been studied considering a static spherically symmetric distribution as well as a spatially homogeneous and isotropic cosmological model.
	The action for a general class of Non Minimally Coupled Scalar Tensor Theory (NMCSTT) \cite{Nordtvedt:1970uv}, \cite{Fujii:2003pa}, \cite{Quiros:2019ktw}, \cite{Faraoni:2004pi}  is given by
	\begin{equation}
		\mathcal{S}=\int \sqrt{-g}~d^{4}x~\left[\mathcal{G}(\Phi) R-\frac{W(\Phi)}{\Phi}\partial_{a}\Phi \partial^{a}\Phi+2\mathcal{L}_{m}\right]\label{eq73*}
	\end{equation} where $\Phi$ is the scalar field. $W$ is an arbitrary function of $\Phi$ in Nordtvedt's formulation.
Varying equation (\ref{eq73*}) w.r.t metric $g_{ab}$ and scalar field $\Phi$ respectively we get the field equations as
\begin{eqnarray}
	\mathcal{G}\left(R_{ab}-\frac{1}{2}Rg_{ab}\right)+g_{ab}\square \mathcal{G}-\nabla_{b}\nabla_{a}\mathcal{G}-\dfrac{W}{\Phi}\Phi \partial^{a}\Phi+\dfrac{1}{2}g_{ab}\dfrac{W}{\Phi}\Phi \partial^{a}\Phi=T_{ab}\label{eq74**}\\
	R\dfrac{d\mathcal{G}}{d\Phi}+\dfrac{2W}{\Phi}\square \Phi+\left(\dfrac{1}{\Phi}\dfrac{dW}{d\Phi}-\dfrac{W}{\Phi^{2}}\right)\Phi \partial^{a}\Phi=0\label{eq75**}
\end{eqnarray}
Thus from equation (\ref{eq74**}) the expression for Raychaudhuri scalar turns out to be
\begin{equation}
	R_{ab}v^{a}v^{b}=\dfrac{1}{2\mathcal{G}}\left[2\left(T_{ab}v^{a}v^{b}+\dfrac{1}{2}T\right)-\square \mathcal{G}+2v^{a}v^{b}\nabla_{b}\nabla_{a}\mathcal{G}+\dfrac{2W}{\Phi}v^{a}v^{b}\partial_{a}\Phi \partial^{a}\Phi\right]\label{eq76**}
\end{equation} The above expression shows that SEC on matter does not necessarily imply the CC in NMCSTT. Further the authors considered a spherically symmetric static space time and investigated the modified CC in Brans-Dicke and Bekenstein's theory. In case of Brans-Dicke theory one has $\mathcal{G}(\Phi)=\Phi$ and $W=$constant. The metric for a static spherically symmetric space time in isotropic coordinates can be written as 
\begin{equation}
	ds^{2}=-C^{2}(r)dt^{2}+D^{^{2}}(r)(dr^{2}+r^{2}d\theta^{2}+r^{2}\sin^{2}\theta d\xi^{2})\label{eq77**}
\end{equation}
For a freely falling observer,
\begin{equation}
	v^{a}=\left(\dfrac{1}{C^{2}},-\dfrac{\sqrt{1-C^{2}}}{CD},0,0\right)
	\end{equation}
It is to be noted that the radial coordinate of velocity is negative that takes into account radial geodesics. In this case we have,\scriptsize
\begin{eqnarray} R_{ab}v^{a}v^{b}=\dfrac{1}{2\Phi}\left[2\left(T_{ab}v^{a}v^{b}+\dfrac{1}{2}T\right)+\dfrac{2-3C^{2}}{C^{2}}\square \Phi+\dfrac{2W}{\Phi}\dfrac{(1-C^{2})}{C^{2}D^{2}}(\dot{\Phi})^{^{2}} -\dfrac{2\dot{\Phi}}{C^{2}D^{2}}\left(2(1-C^{2})\dfrac{\dot{D}}{D}+(2-C^{2})\dfrac{\dot{C}}{C}+\dfrac{2}{r}(1-C^{2})\right)\right]
\end{eqnarray}
\large
If $R_{ab}v^{a}v^{b}=Q_{1}+Q_{2}+Q_{3}+Q_{4}$ then
\begin{eqnarray}
	Q_{1}=\dfrac{1}{\Phi}\left(T_{ab}v^{a}v^{b}+\dfrac{1}{2}T\right)\\
	Q_{2}=\dfrac{3C^{2}-2}{C^{2}}\dfrac{R}{4W}\\
	Q_{3}=\dfrac{(\dot{\Phi})^{2}}{2C^{2}D^{2}\Phi^{2}}\left[(1+2W)-\left(\dfrac{3}{2}+2W\right)C^{2}\right]\\
		Q_{4}=-\dfrac{\dot{\Phi}}{C^{2}D^{2}\Phi}\left[2(1-C^{2})\dfrac{\dot{D}}{D}+(2-C^{2})\dfrac{\dot{C}}{C}+\dfrac{2}{r}\left(1-C^{2}\right)\right]
	\end{eqnarray}
Then considering Brans class I solution $Q_{2}$, $Q_{3}$, $Q_{4}$ and $R_{ab}v^{a}v^{b}$ have been plotted radially for positive and negative $W$. It has been found that avoidance of singularity is guaranteed for negative $W$. In case of Bekenstein's theory one has $\mathcal{G}=1-\dfrac{\Phi^{2}}{6}$ and $W=\Phi$. The metric (Type A solution) is given by
\begin{equation}
	ds^{2}=-C^{2}(r)dt^{2}+D^{2}(r)dr^{2}+S^{2}(r)\left(S^{2}(r)(r^{2}d\theta^{2}+r^{2}\sin^{2}\theta d\xi^{2})\right)
	\end{equation} where
\begin{eqnarray}
	C^{2}(r)=\dfrac{1}{4}\left[W(r)^{\beta}+W(r)^{-\beta}\right]^{2}(W(r))^{2\alpha}\\
	D^{2}(r)=\dfrac{1}{4}\left[W(r)^{\beta}+W(r)^{-\beta}\right]^{2}(W(r))^{-2\alpha}\\
	S^{2}(r)=\dfrac{1}{4}\left[W(r)^{\beta}+W(r)^{-\beta}\right]^{2}(W(r))^{-2\alpha+2}
\end{eqnarray} and $\alpha=\sqrt{1-3\beta^{2}}$, $\beta$ is a constant such that $-\dfrac{1}{\sqrt{3}}\leq \beta \leq \dfrac{1}{\sqrt{3}}$. It reduces to Schwarzschild solution if $W=\sqrt{1-\dfrac{2M}{r}}$ and $\beta=0$. For a freely falling observer,
\begin{equation}
v^{a}=\left(\dfrac{1}{C^{2}},-\dfrac{\sqrt{1-C^{2}}}{CD},0,0\right)
\end{equation} Again writing, $R_{ab}v^{a}v^{b}=R_{1}+R_{2}+R_{3}$ we have
\begin{eqnarray}
	R_{1}=\dfrac{\Phi\square \Phi}{1-\dfrac{\Phi^{2}}{6}}\left(\dfrac{3C^{2}-2}{6C^{2}}\right)\\
	R_{2}=\dfrac{\Phi \dot{\Phi}}{3C^{2}D^{2}\left(1-\dfrac{\Phi^{2}}{6}\right)}\left[(2-C^{2})\dfrac{\dot{C}}{C}+2(1-C^{2})\dfrac{\dot{S}}{S}+\dfrac{2}{r}\left(1-C^{2}\right)\right]\\
	R_{3}=\dfrac{(\dot{\Phi})^{2}}{\left(1-\dfrac{\Phi^{2}}{6}\right)}\left(\dfrac{4-3C^{2}}{6C^{2}D^{2}}\right)
	\end{eqnarray}
The findings from the plot is $R_{ab}v^{a}v^{b}<0$ in the region near $r=2M$. For $\dfrac{1}{2}\leq |\beta|\leq \dfrac{1}{\sqrt{3}}$, $R_{ab}v^{a}v^{b}>0$. CC is violated only for $|\beta|<\dfrac{1}{2}$ for a small range of $r$. For all other choices of $\beta$, CC is always satisfied and a singularity is inevitable. The authors further considered another example in which the background is spatially homogeneous and isotropic FLRW universe given by the metric
\begin{equation}
	ds^{2}=-dt^{2}+a^{2}(t)\left(\dfrac{dr^{2}}{1-kr^{2}}+r^{2}d\theta^{2}+r^{2}\sin^{2}\theta d\xi^{2}\right)
\end{equation} and perfect fluid as matter content of the universe given by
\begin{equation}
	T_{ab}=(\rho+p)v_{a}v_{b}+pg_{ab}
\end{equation} and $v^{a}=(1,0,0,0)$. In case of Brans-Dicke theory (using the field equations (\ref{eq74**}) and \ref{eq75**}) the expression for $R_{ab}v^{a}v^{b}$ is given by 
\begin{equation}R_{ab}v^{a}v^{b}=\dfrac{W\Phi}{2W+3} (3p-\rho)+3\dfrac{\dot{a}^{2}}{a^{2}}+\dfrac{3k}{a^{2}}+\dfrac{W}{2}\dfrac{\dot{\Phi}^{2}}{\Phi^{2}}
	\end{equation}
Thus it can be concluded that in case of flat geometry with pressureless matter $R_{ab}v^{a}v^{b}=\dfrac{6(W+1)(W+2)}{t^{2}(3W+4)^{2}}<0$ for $-2<W-\dfrac{3}{2}$ which resembles with the condition for accelerated expansion of the universe, while violation of CC in this case corresponds to observationally unfavored values of $W$. The analysis shows that CC does not solely depend on the SEC on matter but also on the sign of the parameter $W$. For Bekenstein's scalar field model the field equations yield
\begin{equation}
	R_{ab}v^{a}v^{b}=\dfrac{3p-\rho}{2}+\dfrac{3\dot{a}^{2}}{a^{2}}+\dfrac{3k}{a^{2}}
\end{equation}
With a radiation fluid distribution ($p=\frac{1}{3}\rho$), an open universe may avoid a singularity. Further considering explicit examples of exact solutions containing radiation fluid for $k=0,-1,1$, $R_{ab}v^{a}v^{b}$ have been determined and it was found positive definite for flat and closed universe thus making the focusing inevitable there. However with $k=-1$, $R_{ab}v^{a}v^{b}<0$ and hence there is a possibility of defocusing of geodesics. Thus, avoidance of singularity of proper zero volume might be guaranteed in an open FLRW universe. In this context if we recall the focusing theorem for open universe considering GR, we find that violation of SEC can only avoid focusing and hence singularity formation. In contrary to this if we consider scalar tensor theory of modified gravity then an open universe might avoid singularity even if SEC holds. This is the reason why RE is equally important in modified gravity theories in the context of avoidance of singularity.
	\subsection{Raychaudhuri equation in Kantowski-Sachs space-time model}
	With an aim to formulate the RE in anisotropic background and to find the effect of anisotropy in CC we consider a general metric for an homogeneous and anisotropic space-time with spatial section topology $\mathbf{R\times S^{2}}$. The anisotropy plays a significant role in early stages of evolution of the universe and hence the study of spatially homogeneous and anisotropic cosmological models is physically and cosmologically significant. Observations state that the Universe is homogeneous and isotropic when the inflationary phase was successfully produced (see \cite{Barrow:1987ia}, \cite{Chakraborty:2023neh} for details of inflation). However Cosmic Microwave Background Radiation (CMBR) anomalies \cite{Schwarz:2015cma} concluded that there was an anisotropic phase in the early Universe which does not make it exactly uniform. There lies the motivation of the present work where we consider anisotropic Universe described by Kantowski-Sachs (KS) space-time models and explore the effect/role of anisotropy in RE and CC. 	The exact solutions for homogeneous spacetimes in GR belongs to either Kantowski-Sachs model or the Bianchi Types. KS is the only anisotropic but spatially homogeneous cosmology that does not fall under the Bianchi classification \cite{Leon:2013bra}, \cite{Datta:2021jwr}. These models gained popularity with the publication by Kantowski and Sachs \cite{Kantowski:1966te}. KS space-time has some exciting features. Firstly, its classical and quantum solutions are well known in different contexts \cite{Leon:2010pu}-\cite{Barbosa:2004kp}. Secondly, these models exhibit spherical and translational symmetry and can be treated as non empty analogs of a part of the extended Schwarzschild metric \cite{Collins:1977fg}. Moreover these models help to study the behavior of the added degrees of freedom in quantum cosmological models. Finally, it may be possible that constructing a KS quantum cosmological model may suggest modifications and adaptions in the quantization methods applied to cosmology. Firstly, we present a straightforward derivation of the Raychaudhuri equation in KS model. The Kantowski-Sachs (KS) space-time described by the metric \cite{deCesare:2020swb}
	\begin{equation}
		ds^{2}=-dt^{2}+M^{2}(t)dr^{2}+N^{2}(t)(d\theta^{2}+\sin^{2}\theta d\phi^{2})\label{eq12*}
	\end{equation} where $M(t)$ and $N(t)$ are two arbitrary and independent functions of cosmic time $t$. The generic form of the energy-momentum tensor in support of this geometry is given by
	\begin{equation}
		T^{a}_{b}= diag~(-\rho, p_{r},p_{t}, p_{t})\label{eq12}
	\end{equation} where $\rho$ is the energy density of the physical fluid, $p_{r}$ is the radial and $p_{t}$ is the lateral pressure of the physical fluid. The Einstein's field equations defining the above metric (\ref{eq12*}) and energy-momentum source (\ref{eq12}) can be written as \cite{Xanthopoulos:1992fh}, \cite{Mendes:1990eb}
	\begin{eqnarray}
		\dfrac{\dot{N}^{2}}{N^{2}}+2\left(\dfrac{\dot{M}}{M}\right)\left(\dfrac{\dot{N}}{N}\right)+\dfrac{1}{N^{2}}=\kappa \rho\label{eq98}\\
		2\dfrac{\ddot{N}}{N}+\left(\dfrac{\dot{N}}{N}\right)^{2}+\dfrac{1}{N^{2}}=-\kappa p_{r}\label{eq99}\\
		\dfrac{\ddot{M}}{M}+\dfrac{\ddot{N}}{N}+\left(\dfrac{\dot{M}}{M}\right)\left(\dfrac{\dot{N}}{N}\right)=-\kappa p_{t}\label{eq100}
	\end{eqnarray} where $\kappa=8\pi G$ is the four dimensional gravitational coupling constant and in units $8 \pi G=1$. Further $p_{r}=\omega_{r}\rho$, $p_{t}=\omega_{t}\rho$ and $\omega_{r}\neq \omega_{t}$ i.e, for the sake of generality we consider distinct EoS for the radial and lateral pressures $p_{r}$ and $p_{t}$ respectively. Thus we introduce metric (\ref{eq12*}) and corresponding field equations (\ref{eq98})-(\ref{eq100}) in four dimensions. The average Hubble parameter $H$ in this case is given by
	\begin{equation}
		H=\dfrac{1}{3}\left(\dfrac{\dot{M}}{M}+2\dfrac{\dot{N}}{N}\right)=\dfrac{1}{3}\left(H_{M}+2H_{N}\right)
	\end{equation}
	Anisotropy scalar is given by 
	\begin{equation}
		\sigma=\dfrac{1}{\sqrt{3}}\left(\dfrac{\dot{N}}{N}-\dfrac{\dot{M}}{M}\right)=\dfrac{1}{\sqrt{3}}(H_{N}-H_{M})\label{eq17}
	\end{equation} where $H_{M}=\dfrac{\dot{M}}{M}$ and $H_{b}=\dfrac{\dot{N}}{N}$ . Consider the following transformations:
	\begin{eqnarray}
		V^{3}=MN^{2}\label{eq103}\\
		Z=\dfrac{N}{M}\label{eq104}
	\end{eqnarray}
	Under the above transformation one has the following set of equations
	\begin{eqnarray}
		3\dfrac{\dot{M}}{M}=3\dfrac{\dot{V}}{V}-2\dfrac{\dot{Z}}{Z}\label{eq20}\\
		3\dfrac{\dot{N}}{N}=3\dfrac{\dot{V}}{V}+\dfrac{\dot{Z}}{Z}\label{eq21}\\
		\dfrac{\ddot{M}}{M}+2\dfrac{\ddot{N}}{N}=3\dfrac{\ddot{V}}{V}+\dfrac{2}{3}\dfrac{\dot{Z}^{2}}{Z^{2}}\label{eq22}
	\end{eqnarray}
	From equations (\ref{eq98}), (\ref{eq103}) and (\ref{eq104}) we have
	\begin{equation}
		3\dfrac{\dot{V}^{2}}{V^{2}}-\dfrac{1}{3}\dfrac{\dot{Z}^{2}}{Z^{2}}+\dfrac{1}{V^{2}Z^{\frac{2}{3}}}=\kappa \rho \label{eq23}
	\end{equation} Using definition of $\Theta$ and eq.(\ref{eq103}) one has
	\begin{equation}
		\Theta=3\dfrac{\dot{V}}{V}\label{eq24}
	\end{equation} and using definition of $\sigma$ and eq.(\ref{eq104}) one has
	\begin{equation}
		\sigma=\dfrac{1}{\sqrt{3}}\dfrac{\dot{Z}}{Z}\label{eq25}
	\end{equation}
	From equations (\ref{eq23}), (\ref{eq24}) and (\ref{eq25}) one has
	\begin{equation}
		\dfrac{\Theta^{2}}{3}-\sigma^{2}+\dfrac{1}{V^{2}Z^{\frac{2}{3}}}=\kappa \rho\label{eq26}
	\end{equation}
	The measure of acceleration corresponding to the two scale factors $M(t)$ and $N(t)$ in terms of the transformed variables $V$ and $Z$ are
	\begin{eqnarray}
		\dfrac{\ddot{N}}{N}=\dfrac{\ddot{V}}{V}-\dfrac{2}{9}\dfrac{\dot{Z}^{2}}{Z^{2}}+\dfrac{1}{3}\dfrac{\ddot{Z}}{Z}+\dfrac{2}{3}\left(\dfrac{\dot{V}}{V}\right)\left(\dfrac{\dot{Z}}{Z}\right)\label{eq27}\\
		\dfrac{\ddot{M}}{M}=\dfrac{\ddot{V}}{V}+\dfrac{10}{9}\dfrac{\dot{Z}^{2}}{Z^{2}}-\dfrac{2}{3}\dfrac{\ddot{Z}}{Z}-\dfrac{4}{3}\left(\dfrac{\dot{V}}{V}\right)\left(\dfrac{\dot{Z}}{Z}\right)
	\end{eqnarray} Therefore the field equations (\ref{eq99}) and (\ref{eq100}) in terms of the transformed variables can be written as
	\begin{eqnarray}
		2\dfrac{\ddot{V}}{V}+\dfrac{2}{3}\dfrac{\ddot{Z}}{Z}+\dfrac{\dot{V}^{2}}{V^{2}}+2\dfrac{\dot{V}}{V}\dfrac{\dot{Z}}{Z}-\dfrac{1}{3}\dfrac{\dot{Z}^{2}}{Z^{2}}+\dfrac{1}{V^{2}Z^{\frac{2}{3}}}=-\kappa p_{r}\label{eq29}\\
		2\dfrac{\ddot{V}}{V}-\dfrac{1}{3}\dfrac{\ddot{Z}}{Z}+\dfrac{\dot{V}^{2}}{V^{2}}-\dfrac{\dot{V}}{V}\dfrac{\dot{Z}}{Z}+\dfrac{2}{3}\dfrac{\dot{Z}^{2}}{Z^{2}}=-\kappa p_{t}\label{eq30}
	\end{eqnarray}
	Also,
	\begin{eqnarray}
		\dfrac{\mathrm{d}\Theta}{\mathrm{d}\tau}+\dfrac{\Theta^{2}}{3}=3\dfrac{\ddot{V}}{V}\\
		\dfrac{\mathrm{d}\sigma}{\mathrm{d}\tau}=\dfrac{1}{\sqrt{3}}\left(\dfrac{\ddot{Z}}{Z}-\dfrac{\dot{Z}^{2}}{Z^{2}}\right)
	\end{eqnarray} 
	By some algebraic manipulation with the field equations we get
	\begin{eqnarray}
		\dfrac{\ddot{Z}}{Z}+3\dfrac{\dot{V}}{V}\dfrac{\dot{Z}}{Z}-\dfrac{\dot{Z}^{2}}{Z^{2}}+\dfrac{1}{V^{2}Z^{\frac{2}{3}}}=\kappa(p_{t}-p_{r})\label{eq33}\\
		\dfrac{\ddot{V}}{V}+\dfrac{2}{9}\dfrac{\dot{Z}^{2}}{Z^{2}}=-\dfrac{\kappa}{6}(\rho+p_{r}+2p_{t})\label{eq34}
	\end{eqnarray}
	Thus the evolution equation for expansion ($\Theta$) is given by
	\begin{equation}
		\dfrac{\mathrm{d}\Theta}{\mathrm{d}\tau}+\dfrac{\Theta^{2}}{3}+2\sigma^{2}=-\dfrac{\kappa}{2}(\rho+p_{r}+2p_{t})\label{eq35}
	\end{equation} 
	The evolution equation for shear ($\sigma$) is given by
	\begin{equation}
		\dfrac{\mathrm{d}\sigma}{\mathrm{d}\tau}+\Theta \sigma+\dfrac{\sigma^{2}}{\sqrt{3}}-\dfrac{\Theta^{2}}{3\sqrt{3}}=-\dfrac{\kappa}{\sqrt{3}}(\rho-\frac{1}{3}(p_{t}-p_{r}))\label{eq36}
	\end{equation}
	One may note that the above equations (\ref{eq35}) and (\ref{eq36}) are first order, coupled and non-linear. Since at present one considers the background of anisotropic space-time, so the evolution of anisotropy scalar is given in equation (\ref{eq36}). However, the evolution of expansion scalar $\Theta$ is of central interest in the context of singularity theorems. Mathematically, the evolution equation for expansion scalar is known as Riccati equation (for ref. see a review of Raychaudhuri equations by Kar and Sengupta \cite{Kar:2006ms}). Historically, the Riccati equation or equation (\ref{eq35}) is known as the Raychaudhuri equation.
	
	\subsubsection{Convergence Condition in KS space-time models}
	In order to study the CC in KS space-time we consider the RE in KS space-time given by equation (\ref{eq35}). Let us write
	\begin{equation}
		\dfrac{\mathrm{d}\Theta}{\mathrm{d}\tau}+\dfrac{\Theta^{2}}{3}=-\tilde{R_{c}}\label{eq37}
	\end{equation}
	where
	\begin{equation}
		\tilde{R_{c}}=\\\dfrac{\kappa}{2}(\rho+p_{r}+2p_{t})+2\sigma^{2}\label{eq38}
	\end{equation}
	Here we consider $p_{r}=\omega_{r}\rho$ and $p_{t}=\omega_{t}\rho$, $\omega_{r}$ and $\omega_{t}$ are the radial and transverse EoS parameters. From equation (\ref{eq37}) we have $\tilde{R_{c}}\geq0$ for convergence. Thus the possibilities of focusing are listed as follows:
	\begin{itemize}
		\item $\dfrac{\kappa}{2}(\rho+p_{r}+2p_{t})\geq0$ i.e, matter satisfies SEC. In this case $	\tilde{R_{c}}\geq0$ (follows from equation \ref{eq38}). Since the positive semi definiteness of $\tilde{R_{c}}$ leads to convergence of the bundle of geodesics therefore we name it as Convergence Scalar and we shall use this term in the following sections.
		\item Matter violates SEC i.e, $\dfrac{\kappa}{2}(\rho+p_{r}+2p_{t})<0$ but $2\sigma^{2}>\dfrac{\kappa}{2}|(\rho+p_{r}+2p_{t})|$. 
	\end{itemize}
	Based on the above discussion we conclude that the anisotropy term favors the convergence/ formation of congruence singularity. However this term alone does not lead to CC. If the matter is attractive in nature i.e, $1+\omega_{r}+2\omega_{t}\geq0$ then CC is automatically satisfied. However if the matter is repulsive in nature i.e, $1+\omega_{r}+2\omega_{t}<0$ then anisotropy can not alone lead to CC but needs an extra condition namely $2\sigma^{2}>\dfrac{\kappa}{2}|(\rho+p_{r}+2p_{t})|$. Thus in comparison to FLRW model the anisotropy term may be interpreted as a matter part which is attractive in nature. This interpretation can be justified from Einstein's idea of gravity where geometry and matter are equivalent quantities. Thus to avoid formation of singularity in KS space-time we need $\dfrac{\kappa}{2}(\rho+p_{r}+2p_{t})<0$ and $2\sigma^{2}<\dfrac{\kappa}{2}|(\rho+p_{r}+2p_{t})|$. In other words, to avoid focusing matter cannot be usual in nature. Further violation of SEC must be dominant over the anisotropy term in order to avoid singularity formation. Hence with usual matter singularity is inevitable in KS models. \\  \\
	Thus the KS space-time classically has a past and a future singularity, which can be an anisotropic structure such as a barrel, cigar, a pancake or an isotropic point like structure depending on the initial conditions on anisotropic shear and matter \cite{Ghorani:2021xrs}. Evolution of geodesics terminates at these classical singularities which is identified by the divergence of expansion and shear scalars in the presence of matter (which contributes to the energy density). Existence of singularities pushes GR to the limits of its validity and hence a quantum gravitational treatment which becomes dominant in strong gravity regimes may alleviate the classical singularity. 
	\subsection{Raychaudhuri equation in FLRW like universe with non zero torsion}
	The symmetry present in Riemaannian geometry ensures torsion free spaces. However, one can treat torsion as an independent geometric field, in addition to the metric. As a consequence, there is a possibility to extend the Riemann space-time to Riemann-Cartan spaces. The torsion tensor is denoted by $S^{a}_{bc}$ and is defined by \cite{Chakraborty:2023neh}
	\begin{equation}
		S^{a}_{bc}=\Gamma^{a}_{[bc]}
	\end{equation} 
Metricity condition i.e, $\nabla_{c}~g_{ab}=0$ leads to $
\Gamma^{a}_{bc}=\tilde{\Gamma^{a}_{bc}}+K^{a}_{bc}$. $\tilde{\Gamma^{a}_{bc}}$ defines the Christoffel symbols and $K^{a}_{bc}$ is the contortion tensor defined by
\begin{equation}
	K_{abc}=S_{abc}+S_{bca}+S_{cba}=S_{abc}+2S_{(bc)a}
\end{equation} with $K_{abc}=K_{[ab]c}$. Torsion vector is given by
\begin{equation}
	S_{a}=S^{b}_{ab}=-S^{b}_{ba}
\end{equation} In particular we have
\begin{equation}
	K^{b}_{ab}=2S_{a}
\end{equation} with $K^{b}_{ba}=0$. In space-times with torsion, the field equation is given by
\begin{equation}
	R_{ab}-\dfrac{1}{2}Rg_{ab}=kT_{ab}
\end{equation} which altogether is formally identical to its general relativistic counterpart except the fact that $R_{[ab]}\neq0$ and $T_{[ab]}\neq0$ (i.e, they are anti-symmetric) which are due to the presence of torsion. Using field equations one has \cite{Bose:2020mdm}
\begin{equation}
	S_{abc}=-\dfrac{k}{4}\left(2s_{bca}+g_{ca}s_{b}-g_{ab}s_{c}\right)\label{eq123}
\end{equation} with $s_{abc}=s_{[ab]c}$ and $s_{a}=s^{b}_{ab}$ are the spin tensor and the spin vectors respectively. Trace of (\ref{eq123}) gives $S_{a}=-\dfrac{ks_{a}}{4}$, relating torsion and spin vectors directly. Introducing a time-like 4 velocity vector field $v_{a}$ (so that $v_{a}v^{a}=-1$) facilitates an ($1+(n-1)$) decomposition of the underlying space-time into time and $n-1$ dimensional hyper-surface. The rate of convergence/divergence of a worldline congruence is governed by the Raychaudhuri equation and in the presence of torsion the equation assumes the form \scriptsize
\begin{equation}
	\dfrac{d\Theta}{d\tau}=-\dfrac{\Theta^{2}}{n-1}-R_{(ab)}v^{a}v^{b}-2\sigma^{2}+2\omega^{2}+\nabla_{a}A^{a}+A_{a}A^{a}+\dfrac{2}{n-1}\Theta S_{a}v^{a}-2S_{(ab)c}v^{a}v^{b}A^{c}-2S_{<ab>c}\sigma^{ab}v^{c}+2S_{[ab]c}\omega^{ab}v^{c}
\end{equation}
\large where $A_{a}=\dot{v_{a}}$ is the 4-acceleration vector. Now we explore the RE in FLRW like model with torsion. For this we consider an FLRW type space-time with non-zero torsion and a family of observers along a time-like congruence tangent to the 4-velocity vector field $v_{a}$. Torsion tensor takes the form
\begin{equation}
	S_{abc}=2\Psi h_{a[b~v_{c}]}.
\end{equation}
Due to the homogeneity of the 3-space, $\Psi(t)$ is a scalar function of time only i.e, $\Psi=\Psi(t)$. The torsion vector is 
\begin{equation}
	S_{a}=-3\Psi v_{a}
\end{equation}is purely time-like. \begin{itemize}
\item If $\Psi>0$ then $S_{a}$ and $v_{a}$ are of opposite sign. $S_{a}$ becomes past directed.
\item If $\Psi<0$ then $S_{a}$ and $v_{a}$ are of same sign. In this case $S_{a}$ becomes future directed.
\end{itemize} Therefore the sign of $\Psi$ fixes the relative orientation of the torsion and the 4-velocity vector. The allowed form of contortion tensor in FLRW-like universe is $K_{abc}=4\Psi v_{[ah_{b}]c}$ so that $K^{b}_{ab}=-6\Psi v_{a}=2S_{a}=-K^{b}_{ab}$ and $K^{b}_{ba}=0$ as expected earlier. Cartan field equations lead to
$ks_{abc}=8\Psi h_{c[a v_{b}]}$, the spin tensor and $ks_{a}=12\Psi v_{a}$, the spin vector. Also one can check that $S_{abc}=-\dfrac{1}{4}ks_{cba}$ and $S_{a}=-\dfrac{1}{4}ks_{a}$. To formulate RE in FLRW like universe with torsion, we must have $\sigma_{ab}=0=\omega_{ab}=A_{a}$. Then the general RE reduces to 
\begin{equation}
	\dfrac{d\Theta}{d\tau}=-\dfrac{\Theta^{2}}{3}-\dfrac{k}{2}(\rho+3p)+2\Theta \Psi
\end{equation} where Einstein- Cartan field equations with perfect fluid have been used. The last term on the right hand side of the above equation carries the effect of torsion in convergence/divergence of the time-like congruence tangent to $v_{a}$ vector field depending on the sign of $\Psi$. Let $\Theta$ be the volume scalar with non zero torsion and $\tilde{\Theta}$ is its torsionless counterpart. Then, $\Theta=\tilde{\Theta}+K^{a}_{ba}v^{b}$ where, $K^{b}_{ab}=-6\Psi v_{a}$. Consequently, we have
\begin{equation}
	\Theta=\tilde{\Theta}+6\Psi=3\left(\dfrac{\dot{a(t)}}{a(t)}\right)+6\Psi=3H\left(1+\dfrac{2\Psi}{H}\right).
\end{equation} Here $\dfrac{\dot{a(t)}}{a(t)}=\dfrac{\tilde{\Theta}}{3}=H$ is the Hubble parameter. In FLRW-type cosmologies the RE in the presence of torsion assumes the form
\begin{equation}
	\dfrac{\ddot{a}}{a}=-\dfrac{k}{6}(\rho+3p)-2\dot{\Psi}-2\left(\dfrac{\dot{a(t)}}{a(t)}\right)\Psi
\end{equation}
Thus we may conclude that if matter satisfies SEC i.e, $(\rho+3p)\geq0$
\begin{enumerate}
	\item $\dot{\Psi}>0$ and $\Psi>0$ implies $R_{ab}v^{a}v^{b}>0$ i.e, CC holds and singularity is inevitable analogous to the GR counterpart.
	\item $\dot{\Psi}<0$ and $\Psi<0$ may sometimes imply $R_{ab}v^{a}v^{b}<0$, the violation of CC.
\end{enumerate}
Thus, if the matter content of the universe is usual in nature then a past directed increasing congruence of time-like worldline emerges from the big-bang singularity while for a future directed decreasing congruence of time-like worldline there is no initial singularity. The modified RE in the presence of torsion again hints that CC does not solely depend on the SEC but also on the sign of torsion scalar function.
	\subsection{Implications of Raychaudhuri equation in Bouncing cosmology}
	For Friedmann–Lemaître–Robertson-Walker (FLRW) model with $n=4$,  $\Theta=3H=3\frac{\dot{a}}{a}$ and $\sigma^{2}=0$. Thus the RE (\ref{eq3*}) takes the form
	\begin{equation}
		3\dot{H}=-3H^{2}-R_{ab}v^{a}v^{b}.\label{eq3**}
	\end{equation} In the context of cosmology $\tau$ can be treated as the cosmic time $t$ and `.' represents differentiation w.r.t $t$.
		\section*{$\underline{\mathbf{B1}}$: Bouncing Point is a local minima for $a(t)$}
	\textbf{Behavior of cosmic scale factor $a(t)$ and Hubble parameter $H$:}
	\begin{itemize}
		\item $a(t)$ decreases before bounce, attains a minimum at the bouncing point and then increases after the bounce.
		\item $H<0$ before bounce, $H=0$ at the bouncing point and $H>0$ after the bounce.
		\item Continuity of $H$ clearly depicts that $H$ is an increasing function i.e.  $\dot{H}>0$ in the deleted neighbourhood of bounce.
	\end{itemize}
\textbf{\underline{RE and B1}:} In this case we have $\dot{H}>0$ in the deleted neighbourhood of the bouncing point in $\textbf{B1}$. Therefore from the RE (\ref{eq3**}) one has\\
(i) $R_{ab}v^{a}v^{b}<0$ and, \\
(ii) $|R_{ab}v^{a}v^{b}|>3H^{2}$ in the deleted neighbourhood of $\textbf{B1}$. \\If the matter content of the universe is a perfect fluid having energy density $\rho$ and pressure $p$ then  $R_{ab}v^{a}v^{b}=\dfrac{1}{2}(\rho+3p)$. The RE in FLRW space-time is given by
\begin{equation}
	\dfrac{\ddot{a}}{a}=-\dfrac{4}{3}(\rho+3p)\label{eq5**}
\end{equation} Thus using (i) in equation (\ref{eq5**}) one may conclude that in this bouncing model there is always acceleration. Violation of SEC indicates that matter is exotic in nature and singularity can be avoided near the bounce (in the sense that CC needed for focusing of geodesic congruence is violated here). Further in this type of bounce, the behavior of energy density $\rho$ can be studied using the RE. From the first Friedmann equation we have $3H^{2}=\rho$ and from the RE (\ref{eq3**}) we have $2\dot{H}=-(\rho+p)$ which is nothing but the second Friedmann-equation. Hence for $\textbf{B1}$ in the deleted neighbourhood of bounce $(\rho+p)<0$ i.e. Null Energy Condition (NEC) is violated. To study the continuity of $\rho$ we shall use the matter conservation equation 
\begin{equation}
	\dot{\rho}+3H(p+\rho)=0\label{eq4}
\end{equation} Thus before bounce as $H<0$ so $\dot{\rho}<0$. At bouncing point $H=0$ and hence from the first Friedmann equation $\rho=0$. After bounce $H>0$ and hence $\dot{\rho}>0$. Thus continuity of $\rho$ can be studied as: $\rho$ is decreasing before bounce, attains zero value at bouncing point and increases after bounce. If we assume further that the matter content of the universe is a perfect fluid with barotropic equation of state $p=\omega\rho$, $\omega$ being the equation of state (EoS) parameter then in the neighbourhood of $\textbf{B1}$, $\omega<-1$ i.e. phantom energy favors this type of bounce. From the second Friedmann equation $\dot{H}=0$ at the bouncing point. But $q$, the deceleration parameter is not defined at the bouncing point. Although this type of bounce can avoid the initial big-bang singularity as evident from the RE, yet the peculiarity observed at the bouncing point hints that the bouncing point may be regarded as a higher order singularity in the sense that $a$, $H$, $\dot{H}$ are defined at the bouncing point but no other higher order parameters like $q$ and $j$ are defined at the bouncing point. However there is always acceleration in this type of model except at the bouncing point where $\ddot{a}=0$.
	\section*{$\underline{\mathbf{B2}}$ : Bouncing Point is a local maxima for $a(t)$}
\textbf{Behavior of cosmic scale factor $a(t)$ and Hubble parameter $H$:}
\begin{itemize}
	\item $a(t)$ increases before bounce, attains a maximum at the bouncing point and then decreases after the bounce. 
	\item $H>0$ before bounce, $H=0$ at the bouncing point and $H<0$ after the bounce.
	\item Continuity of $H$  clearly depicts that $H$ is a decreasing function i.e.  $\dot{H}<0$ in the deleted neighbourhood of bounce.
\end{itemize}
\textbf{\underline{RE and B2}:} For $\textbf{B2}$, $\dot{H}<0$ in the  neighbourhood of the bouncing point and $H=0$ at the bouncing point. Therefore from the RE (\ref{eq3**}) there are two possibilities: (i) $R_{ab}v^{a}v^{b}>0$ or (ii) $R_{ab}v^{a}v^{b}<0$ but $|R_{ab}v^{a}v^{b}|<3H^{2}$ in the neighbourhood of the bouncing point in $\textbf{B2}$. 
\begin{enumerate}
	\item In the first case convergence condition (CC) or precisely the Strong Energy Condition (SEC) holds. Therefore bounce occurs even with normal/usual matter. There is always deceleration except at the bouncing epoch as clear from the RE (\ref{eq5**}). Since $a_{b}\neq0$, hence at the bouncing point there is non-zero volume (hence no singularity). However focusing/convergence condition holds in the neighborhood of bounce. This shows focusing alone does not always leads to singularity formation. However the converse is true i.e, if there is a singularity of the space-time then a congruence of time-like/ null geodesic will focus there.
	\item  For the second case bounce occurs with exotic matter having EoS parameter $-1<\omega<-\frac{1}{3}$ and violation of CC results in the avoidance of big-bang singularity. There is always acceleration except at the bouncing point where $\ddot{a}=0$.
\end{enumerate} The energy density $\rho$ shows a similar behavior just like in $\textbf{B1}$. $\dot{H}=0$ at the bouncing point. In $\textbf{B2}$ also $a$, $H$ and $\dot{H}$ are defined at the bouncing point but next higher order derivatives are undefined at the bouncing point. This absurdness of the bouncing point again hints the existence of 2nd order singularity. Therefore the RE essentially depicts the existence of higher order singularities at the bouncing point. In this context the paper \cite{Chakraborty:2023lav} shows an extensive analysis of the bouncing cosmology via the celebrated RE. The geometry of the bouncing point was explored and used as a tool to classify the models from the point of view of cosmology and RE was furnished in various bouncing models to classify the bouncing points as regular point or singular point. Also RE hints various issues with these two kinds of bounce for example, the theoretical implications of $\textbf{B1}$ demands to break a series of singularity theorems by Hawking and Penrose which uses RE as a key ingredient. Such an issue is accompanied by violation of NEC since we restricted our study within the framework of GR. Thus construction of $\textbf{B1}$ in reality without theoretical pathologies is not easy as the scenario is associated with NEC violation which is again accompanied by quantum instabilities. The cosmological model involving a contraction phase (e.g $\textbf{B1}$) suffers from BKL (Belinsky-Khalatinov-Lifschitz) instability issue. Also the examples of $\textbf{B1}$ are non-singular bouncing model. This means they eradicate the singularity by constructing a universe that begins with a contracting phase and then bounces back to an expanding phase.  After years of continuous efforts, it is proposed that an effective field theoretic description combining the benefits of matter bounce and Ekpyrotic scenarios can give rise to a non singular cosmological model without pathologies through a Galileon-like Lagrangian. For RE in $\textbf{B2}$ there are two possibilities. In the first case, occurrence of bounce is realistic with usual matter. For the second case bounce occurs with exotic matter.
\subsection{Raychaudhuri equation and Emergent Universe}
The RE for a hyper-surface orthogonal congruence of time-like geodesic is given by
\begin{equation}
	\dfrac{d\Theta}{d\tau}=-\dfrac{\Theta^{2}}{3}-2\sigma^{2}-\tilde{R}
\end{equation} where $\tilde{R}=R_{ab}v^{a}v^{b}$. In FLRW (homogeneous and isotropic) case, $\Theta=3H$, $H=\dfrac{\dot{a}}{a}$ (Hubble parameter) and due to isotropy $2\sigma^{2}=0$. In cosmology, $\tau$ is nothing but the cosmic time $t$. So RE in terms of the cosmic parameter $H$ and cosmic time $t$ takes the form
\begin{equation}
	\dot{H}+H^{2}=-\dfrac{1}{3}\tilde{R}
\end{equation} which in terms of scale factor can be written as
\begin{equation}
	\dfrac{\ddot{a}}{a}=-\dfrac{1}{3}\tilde{R}
\end{equation} Previously we have learned that $\tilde{R}\geq0$ is the CC. In order to examine the sign of Raychaudhuri scalar for emergent scenario we consider the following asymptotic behavior for the Hubble parameter and scale factor as \cite{Haldar:2017tqd}, \cite{Chakraborty:2014ora}
\begin{enumerate}
	\item $a\rightarrow a_{0}$, $H\rightarrow0$ as $t\rightarrow -\infty$.
	\item $a \approx a_{0}$, $H \approx 0$ for $t\leq t_{0}$.
	\item $a=a_{0}e^{H_{0}(t-t_{0})}$, $H\approx H_{0}$ for  $t\leq t_{0}$.
\end{enumerate} Here $a_{0}$, a constant is the tiny value of the scale factor during emergent era of evolution and $H_{0}>0$ is the value of the Hubble parameter at $t=t_{0}$. Thus from the RE, employing the above conditions we can find the nature of $\tilde{R}$ in emergent scenario. This is because it is the sign of $\tilde{R}$ that plays a significant role in Focusing theorem. We find that:
\begin{enumerate}
	\item as $t\rightarrow -\infty$ (infinite past), $\tilde{R}\rightarrow0$
	\item For $t\leq t_{0}$, $\tilde{R}\approx -3H_{0}^{2}<0$
\end{enumerate} Thus throughout the time axis $(-\infty,+\infty)$ we get $\tilde{R}\leq0$. Thus CC is violated. This leads to the avoidance of singularity. This is why, emergent universe is a non-singular model of universe where the initial big-bang singularity is replaced by an Einstein static era with $a=a_{0}\neq 0$. In the emergent era RE leads to the violation of Focusing theorem as $\tilde{R}\leq0\implies 	\dfrac{d\Theta}{d\tau}+\dfrac{\Theta^{2}}{3}\geq0.$ This allows the defocusing of geodesics thereby guaranteeing the singularity free nature of emergent universe. This is a case which shows that in a singularity free model, geodesics are complete and there is no focusing. The formal definition of singularity first appeared in the works of Penrose where he defined singularity from the point of view of geodesic incompleteness. Thus, from the definition itself it is also clear that a singularity free model has complete geodesics.
	\subsection{Role of Raychaudhuri equation in Cosmic Evolution}
	The Raychaudhuri equation for hyper-surface orthogonal congruence of time-like geodesics in $(n+1)$-dimensional homogeneous and isotropic FLRW space-time is given by equation
	\begin{equation}
		\dfrac{d\Theta}{d\tau}=-\dfrac{\Theta^{2}}{n}-\tilde{R}\label{eq139}
	\end{equation}
	In order to study the role of RE in cosmic evolution we consider the following transformation given by
	\begin{equation}
		X=\sqrt{\eta}=a^{3},\label{eq140}
	\end{equation}
	where $\eta$=det($\eta_{ab}$) is the determinant of the metric of the n-dimensional space-like hyper-surface provided the space time manifold is of dimension $n+1$. The dynamical evolution of $\eta$ is given by,
	\begin{equation}
		\dfrac{1}{\sqrt{\eta}}\dfrac{d\sqrt{\eta}}{d\tau}=\Theta,
	\end{equation} so that \begin{equation}
		\dfrac{dX}{d\tau}=X\Theta.\label{eq142}\end{equation}
	Hence for ($n$+1)-dimensional space-time manifold the RE can be written as a second order nonlinear ordinary differential equation as,
	\begin{equation}
		\dfrac{X''}{X}+ \left(\dfrac{1}{n}-1\right)\left(\dfrac{X'}{X}\right)^{2}+\tilde{R}=0.\label{eq143}
	\end{equation} ` $'$ ' denotes differentiation w.r.t $\tau$. The above second order non-linear differential equation has a first integral of the form 
	\begin{equation}\label{eq22***}
		H^{2}=\dfrac{a^{-\frac{6}{n}}}{9}\left[u_{0}-6\int a^{(\frac{6}{n}-1)}\tilde{R}~da\right],
	\end{equation} where $H=\dfrac{\dot{a}}{a}$ is the Hubble parameter, $a(t)$ is the scale factor and $u_{0}$ is a constant of integration.\\
	The above first integral for 4-D space-time ($n=3)$ can be identified as the first Friedmann equation
	\begin{equation} 3H^{2}+\dfrac{\kappa}{a^{2}}=\rho\nonumber
		\end{equation}with 
	\begin{equation}
		\kappa=-\dfrac{u_{0}}{3}\label{eq23**}
	\end{equation}
	and,
	\begin{equation}
		\rho=-\dfrac{2}{a^{2}}\int a~\tilde{R}~da\label{eq24**}.
	\end{equation} Equation (\ref{eq23**}) hints that $u_{0}$ is not merely a constant of integration but is related to the geometry of space-time as 
	$u_{0}>=<0$ for open/flat/closed model. Further one may show that (\ref{eq24**}) holds in Einstein gravity as a particular case. It is to be noted that from the first integral (\ref{eq22***}) one can find the cosmological solutions or the scenario of cosmic evolution as follows:\\
	 If matter is in the form of perfect fluid with equation of state $p=\omega(a)\rho$, then energy momentum conservation equation $\dot{\rho}+3\rho(1+\omega(a))H=0$ gives 
\begin{equation}\label{eq26***}
	H^{2}=\dfrac{a^{-\frac{6}{n}}}{9}\left[u_{0}-3\rho_{0}\int a^{(\frac{6}{n}-4)}~(1+3\omega(a))~\exp\left(-3\int \dfrac{w(a)}{a} da\right)da\right].
\end{equation} Therefore for dust era of evolution one has
\begin{equation}
	H^{2}=\dfrac{u_{0}}{9} a^{\frac{-6}{n}}-\dfrac{\rho_{0}a^{-3}}{3(\frac{6}{n}-3)}.
\end{equation} For 4D- space-time $n=3$, so
\begin{equation}
	H^{2}=\dfrac{u_{0}}{9}a^{-2}+\dfrac{\rho_{0}}{3}a^{-3}.
\end{equation}
Using $H=\dfrac{\dot{a}}{a}$ one has the solution as,
\begin{equation}
	(t-t_{0})=3\int\dfrac{\sqrt{a}}{\sqrt{(3\rho_{0}+u_{0}a)}}da
\end{equation}or,
\begin{equation}
	(t-t_{0})=\dfrac{6}{u_{0}^{\frac{3}{2}}}\left[\dfrac{au_{0}}{2}\sqrt{(3\rho_{0}+au_{0})}-\dfrac{3\rho_{0}}{2}\cosh^{-1}\left(\sqrt{1+\frac{u_{0}a}{3\rho_{0}}}\right)+k\right],
\end{equation} where $k$ is the constant of integration.
When $\omega(a)=\omega_{0},$ a non zero constant then
\begin{equation}
	H^{2}=\dfrac{a^{-\frac{6}{n}}}{9}\left[u_{0}-3\rho_{0}(1+3\omega_{0})\dfrac{a^{(\frac{6}{n}-3-3\omega_{0})}}{(\frac{6}{n}-3-3\omega_{0})}\right].
\end{equation}
For 4-D space-time again putting $n=3$, one gets
\begin{equation}
	H^{2}=\dfrac{u_{0}}{9}a^{-2}+\dfrac{\rho_{0}}{3}a^{-3(1+\omega_{0})},
\end{equation}or
\begin{equation}
	(t-t_{0})=3\int\dfrac{da}{\sqrt{u_{0}+3\rho_{0}a^{-(1+3\omega_{0})}}}.
\end{equation}
The solution for various non-zero choices (of course that make the integration solvable) of $\omega_{0}$ are given below in table (\ref{t1}). 
\begin{table}
\centering	\begin{tabular}{ |p{2cm}|p{9cm}|}
		\hline
		\multicolumn{2}{|c|}{Solution for various non-zero choices of $\omega_{0}$} \\
		\hline
		$\omega_{0}$& Solution  \\
		\hline
		$~~\frac{1}{3}$  &  $a(t)=\left[\dfrac{(t-t_{0})^{2}u_{0}}{9}-\dfrac{3\rho_{0}}{u_{0}}\right]^{\frac{1}{2}}$   \\
		$~~1$& $(t-t_{0})=\dfrac{a^{3}~2^{F_{1}}~\left(\frac{1}{2},\frac{3}{4};\frac{7}{4};\frac{-u_{0}}{3\rho_{0}}a^{4}\right)}{\sqrt{3\rho_{0}}}$  \\
		$-\frac{1}{3}$ & 
		$a(t)=\dfrac{\sqrt{u_{0}+3\rho_{0}}}{3}(t-t_{0})$ \\
		$-1$ & $a(t)=\sqrt{\frac{u_{0}}{3\rho_{0}}}\left[\coth^{2}(\sqrt{\frac{\rho_{0}}{3}}(t-t_{0}))-1\right]^{\frac{-1}{2}}$\\
		\hline
	\end{tabular}\label{t1}
\caption{Cosmic scale factor for various non zero choices of $\omega=\omega_{0}$}
\end{table}
The solution for a general $w(a)$ of the form [ for ref. see  	\cite{Jassal:2005qc}] \begin{equation} 
	\omega(a)=\omega_{0}+\omega^{'}\left(\dfrac{a}{a-1}\right),\end{equation}is given by
\begin{equation}
	\dot{a}=\dfrac{1}{3}\left[u_{0}-3\rho_{0}\int((1+3\omega_{0})(a-1)+3\omega^{'}a)(a-1)^{-1-3\omega^{'}}a^{-(2+3\omega_{0})}da\right]^{\frac{1}{2}}
\end{equation} which upon further simplification yields
\begin{equation}
	3\int\dfrac{da}{\left[u_{0}-3\rho_{0}~A\right]^{\frac{1}{2}}}=(t-t_{0}),
\end{equation} where $A$ is given by
\begin{equation}
	\scriptsize	A=\dfrac{(a-1)^{-3\omega^{'}}}{3\omega^{'}}\left[-(1+3\omega_{0}+3\omega^{'})~2^{F_{1}}(1+3\omega_{0},-3\omega^{'};1-3\omega^{'};1-a)+(1+3\omega_{0})~2^{F_{1}}~(2+3\omega_{0},-3\omega^{'};1-3\omega^{'};1-a)\right].
\end{equation}
and $2^{F_{1}}$ is the Gauss-Hypergeometric function.\\ In case of inflation $\tilde{R}=\dfrac{1}{2}(\rho_{\phi}+3p_{\phi})\simeq-V_{0}$. So one has
\begin{equation}
	H^{2}=\dfrac{nV_{0}}{9}+\dfrac{u_{0}}{9}a^{-\frac{6}{n}}.
\end{equation}
For 4-D space-time the solution becomes,
\begin{equation}
	(t-t_{0})=\int\dfrac{da}{\sqrt{\frac{u_{0}}{9}+\frac{V_{0}a^{2}}{3}}},
\end{equation}or
\begin{equation}
	(t-t_{0})=\sqrt{\dfrac{3}{V_{0}}}~\ln|a|+\left(a^{2}+\frac{u_{0}}{3V_{0}}\right)^{\frac{1}{2}}+C,
\end{equation}\\ \\ $C$, being the constant of integration. Thus a detailed analysis of these solutions can be made graphically using the first integral constructed out of a second order differential equation in which the first order RE was converted to can be done and compared with the latest observational dataset. By choosing perfect fluid with barotropic equation of state as the matter content of the universe, cosmological solutions can be obtained for various choices of the parameters in the equation of state. 
	\section{Raychaudhuri equation from the point of view of Mechanics}
	\subsection{Harmonic Oscillator}
	From mathematical point of view the RE (\ref{eq2**}) can be termed as Riccati equation and it becomes a linear second order equation as \cite{Chakraborty:2023ork}
	\begin{equation}
		\dfrac{d^{2}Y}{d\tau^{2}}+\omega_{0}^{2}~Y=0,\label{eq8}
	\end{equation} where $\Theta=(n-1)\dfrac{d}{d\tau}\ln Y$, $\omega_{0}^{2}=\dfrac{1}{n-1}(\tilde{R}+2\sigma^{2}-2\omega^{2}-\nabla_{b}A^{b}).$ Thus the RE can be identified as a linear harmonic oscillator equation with time varying frequency $\omega_{0}$. As 	$\Theta$ may be defined as the derivative of the geometric entropy ($S$) or an average (or effective)
	geodesic deviation so one may identify $S = \ln Y$. The expansion $\Theta$ is nothing but the rate of change of volume of the transverse subspace of the congruence/bundle of geodesics. Therefore, the expansion approaching negative infinity (i.e. $\Theta\rightarrow-\infty$) implies a convergence of
	the bundle, whereas a value of positive infinity (i.e. $\Theta\rightarrow+\infty$) would imply a complete divergence. Thus the
	Convergence Condition (CC) can be stated as follows:\\ \\
	$~~~~$	(i) Initially $Y$ is positive but decreases with proper time i.e $\dfrac{dY}{d\tau}<0$.\\

	(ii)  Subsequently $Y=0$ at a finite proper time to have negative infinite expansion.\\ \\
	From the above interrelation: $\Theta=\dfrac{(n-1)}{Y}\dfrac{dY}{d\tau}$, it is clear that there should be an initially negative expansion (i.e. $\Theta(\tau=0)<0$) and subsequently $\Theta\rightarrow-\infty$ as $Y\rightarrow0$ at a finite proper time. Therefore the CC essentially coincides with the condition for the existence of zeroes of $Y$ in finite proper time. However the Sturm comparison theorem (in differential equation) shows that the existence of zeros in $Y$ at finite value of the proper time $\tau$ requires
	\begin{equation}
		(\tilde{R}+2\sigma^{2}-2\omega^{2}-\nabla_{c}A^{c})\geq0\label{eq34*}.
	\end{equation} Therefore (\ref{eq34*}) is the CC for a congruence of time-like curves (may be geodesic or non geodesic). Further, the above inequality shows that, the Raychaudhuri scalar $\tilde{R}$ and the shear/anisotropy scalar $2\sigma^{2}$ are in favor of convergence of the congruence of time-like curves while rotation and acceleration terms oppose the convergence. The CC reduces to $\tilde{R}\geq0$ if we consider the congruence of time-like
	curves to be geodesic and orthogonal to the space-like hyper-surface. This leads to \textbf{Geodesic
		Focusing} and hence the \textbf{Focusing Theorem}. In other words, rotation and acceleration terms act against the focusing but shear and
	Raychaudhuri scalar are in favor of it. Thus from physical point of view, if the RE corresponds to a realistic linear harmonic oscillator then it is inevitable to have a singularity. We name the scalar: $R_{c}=\tilde{R}+2\sigma^{2}$, as the Convergence scalar.
	Since in other modified theories of gravity the field equations are different, there may arise certain conditions for the possible avoidance of singularity even with the assumption of Strong Energy Condition (SEC) \cite{Chakraborty:2023ork}. Therefore in the context of avoidance of classical singularity, the role of RE is also equally important in modified gravity theories as discussed earlier\cite{Choudhury:2021zij}.  \\ \\
	Further in the homogeneous and isotropic FLRW background ($\sigma^{2}=0$), the Convergence scalar ($R_{c}$) coincides with the Curvature scalar ($\tilde{R}$) and $\tilde{R}$ can be treated as the Raychaudhuri scalar if we consider a congruence of time-like geodesics ($\nabla_{b}A^{b}=0$) which are hyper-surface orthogonal ($\omega=0$) and moving in a four dimensional space time ($n=4$). As already mentioned in the work of \cite{Albareti:2012se}, \cite{Albareti:2014dxa} $\tilde{R}$ can be treated as the mean curvature geometrically. Thus we attempt to find the solution of equation (\ref{eq8}) in FLRW background assuming some possible forms of $\tilde{R}$. Since the background is homogeneous, $\tilde{R}$ may be treated as a function of the scale factor $a$ say $\dfrac{\tilde{R}}{3}=G(a)$. Thus eq (\ref{eq8}) takes the form
	\begin{equation}
		\dfrac{d^{2}a}{dt^{2}}+G(a)a=0.
	\end{equation}
	Now we find the solution of the above differential equation considering the following cases.\\ \\
	\textbf{Case-I:} $G(a)=l;~l>0$ i.e, we consider a positive constant mean curvature. The solution is given by 
	\begin{equation}
		a(t)=A~\sin(\sqrt{l}t)+~B~\cos(\sqrt{l}t).
	\end{equation}
\textbf{Case-II:} $G(a)=-p~;p>0$ i.e, we consider constant negative mean curvature. In this case the solution is given by 
\begin{equation}
	a(t)=c~\sinh(\sqrt{p}t)+d~\cosh(\sqrt{p}t).
\end{equation}
\textbf{Case-III:} $G(a)=G_{0}a^{n}$. In this case the solution assumes the form
\begin{equation}
	a(t)=-\dfrac{G_{0}n^{2}}{2(n+2)}t^{-\frac{2}{n}}
\end{equation} 
In case of positive constant curvature we get bouncing scale factor (there is an epoch with respect to which there are two phases- expanding and contracting phase). In this case although the CC holds yet there is no curvature singularity. This is in agreement with the result of Landau and Lifschitz \cite{book} which reveals that CC or focusing alone can not imply singularity.
In case of negative constant mean curvature, the CC is violated. The graph of $a(t)$ shows that there is no singularity. Thus in this case  avoidance of geodesic focusing or violation of the CC leads to possible avoidance of singularity.
 In case of variable mean curvature (positive definite) CC holds. This shows that at a strong curvature singularity, the gravitational tidal forces linked with the singularity are so strong that any particle trying to cross it must be crushed to zero size.
	\subsection{Lagrangian and Hamiltonian formulation}
	In the earlier sections we saw that RE, a first order differential equation in $\Theta$ in eq. (\ref{eq139}) can be converted to second order differential equation in a transformed variable $X$ eq.(\ref{eq143}) using eq. (\ref{eq140}). The second order differential equation can be written as
	\begin{equation}
		X\ddot{X}+\left(\frac{1}{n}-1\right)\dot{X}^{2}+(2\sigma^{2}+\tilde{R})X^{2}=0\label{eq168}
	\end{equation} for a general space-time with hyper surface orthogonal congruence of time-like geodesics. Therefore, it is a natural search for a Lagrangian corresponding to which the Euler-Lagrange equation gives (\ref{eq168}). According to Helmholtz \cite{Davis:1928},\cite{Davis:1929},\cite{Douglas:1941},\cite{Casetta:1941},\cite{Crampin:2010},\cite{Nigam:2016}, for a system of 'r' second order differential equations of the form
	\begin{equation}
		\mathcal{\mu}_{\alpha}(\tau,y_{\delta},\dot{y_{\delta}},\ddot{y_{\delta}})=0 ,~~ \alpha,\delta=1,2,...,r
	\end{equation}
	( `.' indicates differentiation w.r.t proper time $\tau$), the necessary and sufficient conditions for being the Euler-Lagrange equations corresponding to a Lagrangian $L(\tau,y_{\delta},\dot{y_{\delta}})$, termed as Helmholtz conditions \cite{Davis:1928}-\cite{Nigam:2016} are given by
	\begin{eqnarray}
		\dfrac{\partial\mu_{\alpha}}{\partial\ddot{y_{\delta}}}=\dfrac{\partial\mu_{\delta}}{\partial\ddot{y_{\alpha}}}\\
		~~~~~~~~~	\dfrac{\partial\mu_{\alpha}}{\partial y_{\delta}}-\dfrac{\partial\mu_{\delta}}{\partial y_{\alpha}}=~\dfrac{1}{2}~\dfrac{d}{d\tau}\left(\dfrac{\partial\mu_{\alpha}}{\partial{\dot{y_{\delta}}}}-\dfrac{\partial{\mu_{\delta}}}{\partial{\dot{y_{\alpha}}}}\right)
	\end{eqnarray} and
	\begin{equation}
		~~~~~~~~~~~~~~~	\dfrac{\partial\mu_{\alpha}}{\partial{\dot{y_{\delta}}}}+\dfrac{\partial{\mu_{\delta}}}{\partial{\dot{y_{\alpha}}}}=2~\dfrac{d}{d\tau}~\left(\dfrac{\partial\mu_{\delta}}{\partial\ddot{y_{\alpha}}}\right)
	\end{equation} with $(\alpha,\delta)=1,2,...,r$.
	In the present context, we have a single second order differential equation (\ref{eq168}) so the above conditions reduce to
	\begin{equation}\label{eq13}
		\dfrac{d\mu}{d\dot{X}}=\dfrac{d}{d\tau}\left(\dfrac{d\mu}{d\ddot{X}}\right)
	\end{equation} with,
	\begin{equation}\label{eq14} \mu(\tau,X,\dot{X},\ddot{X})=X\ddot{X}+\left(\frac{1}{n}-1\right)\dot{X}^{2}+(2\sigma^{2}+\tilde{R})X^{2}
	\end{equation}
	A simple algebraic calculation shows that equation (\ref{eq13}) is satisfied for $\mu$ given in equation (\ref{eq14}) only for $n=\frac{2}{3}$, which is not possible as $n$ is the dimension of the hyper-surface. (If one chooses,  $\mu=\dfrac{\ddot{X}}{X}+\left(\frac{1}{n}-1\right)\dfrac{\dot{X}^{2}}{X^{2}}+(2\sigma^{2}+\tilde{R})$ then (\ref{eq13}) implies $n=2$). Thus for general '$n$', ~ (\ref{eq13}) will be satisfied for $\tilde{\mu}$ , provided
	$\tilde{\mu}=X^{\alpha}\mu$ with $\alpha=\left(\dfrac{2}{n}-3\right).$
	Therefore, one has
	\begin{eqnarray}
		\tilde{\mu}=X^{2\left(\frac{1}{n}-1\right)}\ddot{X}+\left(\frac{1}{n}-1\right)X^{\left(\frac{2}{n}-3\right)}\dot{X}^{2}+(2\sigma^{2}+\tilde{R})X^{\left(\frac{2}{n}-1\right)}\nonumber\\
		=\dfrac{d}{d\tau}\left[X^{2(\frac{1}{n}-1)}\dot{X}\right]-\left(\frac{1}{n}-1\right)X^{(\frac{2}{n}-3)}\dot{X}^{2}+h(X)X^{(\frac{2}{n}-1)}
	\end{eqnarray} provided $(2\sigma^{2}+\tilde{R})$ is  a function of $X$ alone, say $h(X)$.
	Thus one may construct the Lagrangian as
	\begin{equation}
		\mathcal{L}=\dfrac{1}{2}X^{2\left(\frac{1}{n}-1\right)}\dot{X}^{2}-V[X]\label{eq15}
	\end{equation} with,
	\begin{equation}
		\dfrac{\delta V[X]}{\delta X}=X^{(\frac{2}{n}-1)}h(X)\label{eq177}
	\end{equation}
	Now the momentum conjugate to the variable '$X$' is
	\begin{equation}\label{eq18}
		\Pi_{X}=\dfrac{\partial\mathcal{L}}{\partial\dot{X}}=X^{2(\frac{1}{n}-1)}\dot{X}
	\end{equation}
	At this point one may check that the Euler Lagrange equation corresponding to the Lagrangian in eq. (\ref{eq15}) gives back the RE in eq. (\ref{eq168}).
	So, the Hamiltonian of the system is given by
	\begin{equation}
		\mathcal{H}=\dfrac{1}{2}X^{-2\left(\frac{1}{n}-1\right)}\Pi_{X}^{2}+V[X]
	\end{equation}
	One may note that one of the Hamilton's equation of motion gives the Raychaudhuri equation (\ref{eq168}) while the other one yields the definition of momentum (\ref{eq18}).
	From the above formulation of $\tilde{\mu}$, it is clear that $\tilde{\mu}$ satisfies all the Helmholtz conditions provided $2\sigma^{2}+\tilde{R}$ is a sole function of $X$. Now the RE for $n+1$-dimensional space time is given by
	\begin{equation}
		\dfrac{d\Theta}{d\tau}=-\dfrac{\Theta^{2}}{n}-2\sigma^{2}-\tilde{R}
	\end{equation} We denote the scalar $2\sigma^2+\tilde{R}$ by $R_c$. Therefore RE can be written as
	\begin{equation}
		\dfrac{d\Theta}{d\tau}+\dfrac{\Theta^2}{n}=-R_c
	\end{equation} Again since $R_{c}=h(X)$, eq. (\ref{eq177}) shows that $R_{c}$ is gradient of potential and is therefore indicative of force.  $R_c>(<0)$ implies force is attractive (repulsive) in nature. This hints that convergence will occur (i.e. $R_c>0$) if the matter is attractive. This is the reason why RE is regarded as the fundamental equation of gravitational attraction. 
	\section{Raychaudhuri equation in quantum settings}
	After the phenomenal detection of gravitational waves, although Einstein's General Theory of relativity is the most successful theory to describe physical reality, the appearance of singularity in GR is the biggest drawback of the theory itself. Resolution of these singularities is thus a need of the hour to describe the space-time without any peculiarity. According to a general speculation, quantum effects that become dominant in strong gravity regime may alleviate this singularity problem. Therefore there is a natural search for quantum mechanical tools for the possible avoidance of this singularity that persists in a classical space-time at the outset of GR. In quantum correction of the RE, some extra terms are added with ($-R_{c}$) in the R.H.S of the classical RE so that they act as repulsive force to prevent focusing. Appearance of singularity implies convergence/focusing, hence if one can prevent focusing using this quantum correction of RE avoidance of singularity might be guaranteed. (for ref see \cite{Das:2013oda}-\cite{Blanchette:2020kkk}). In particular a quantum version of the Raychaudhuri equation may probably be useful in the context of identifying the existence of a singularity in the quantum regime. Although there is no universally accepted theory of quantum gravity \cite{Thiemann:2007pyv}, there are at present two major approaches for formulating a quantum theory of gravity-- canonical quantization \cite{DeWitt:1967yk} and path integral formulation \cite{Hawking:1978jz}. In canonical quantization , the operator version of the Hamiltonian constraint (known as Wheeler-Dewitt (WD) equation) is a second order hyperbolic functional differential equation and its solution is known as the wave function of the universe \cite{Hartle:2022ykc}. However, it is hard to find a solution of the WD equation even in simple minisuperspace models. Also there is an ambiguity in operator ordering and how to know the initial conditions of the universe to have a well defined wave function. However an important feature of the Hamiltonian in the operator version is that it admits a self adjoint extension in a general sense. As a result, the conservation of probability is ensured. On the other hand, the path integral formulation is more favourable due to some definite proposals for the sum over histories (namely by Hartle \cite{Hartle:2022ykc}, Hawking \cite{Hawking:1978jz} and by Vilenkin \cite{Vilenkin:1988yd}). In the present study we review few quantum mechanical tools. The first way is the Lagrangian and the Hamiltonian formulation for the RE which paves the way for canonical quantization and formulation of WD equation. Further we argue that the solution of the WD equation plays a pivotal role for singularity analysis in quantum regime. This is an attempt to quantize the geodesic flow and avoid singularity in homogeneous cosmology. This formulation is expected to find application in gravitational collapse of homogeneous systems.  While the other way replaces the classical geodesics by quantum Bohmian trajectories which are found to obviate the Big Bang singularity in the presence of non-zero quantum potential. Raychaudhuri equation has been furnished in loop quantum gravity (LQG) by \cite{Blanchette:2020kkk}. They have shown that LQG corrections to the interior of Schwarzschild black hole induce additional terms in the right hand side of the RE. These terms are repulsive in nature near the classical singularity. The quantum contributed terms oppose the convergence prevalent in classical GR. Under these circumstances singularity theorems fail to hold. Consequently, geodesics are no longer incomplete and there is resolution of the singularity that exists at the classical level. Singularity analysis may be carried out for other black hole space-times like RN, Kerr etc using LQG corrections.
	\subsection{Wheeler-DeWitt Quantization: Quantum cosmology}
	For canonical quantization, $X$ and $\Pi_{X}$ are considered as operators acting on the state vector $\mathbf{\Psi}(X,\tau)$ of the geometric flow. In $X$- representation, the operators assume the form $\tilde{X}\rightarrow X$ and $\tilde{\Pi_{X}}\rightarrow -i\hbar \dfrac{\partial}{\partial X}$ so that $[\tilde{X}, \tilde{\Pi_{X}}]=i\hbar$ and the operator form of the Hamiltonian is \cite{Choudhury:2021huy}
	\begin{equation}
		\tilde{\mathcal{H}}=-\dfrac{\hbar^{2}}{2} X^{2(1-\frac{1}{n})} \dfrac{d^{2}}{dX^{2}}+V[X].
	\end{equation} In the context of cosmology, there is notion of Hamiltonian constraint and operator version of it acting on the wave function of the universe ($\mathbf{\Psi}$) gives $\tilde{\mathcal{H}}\mathbf{\Psi}=0$. This is known as the Wheeler-Dewitt (WD) equation which explicitly takes the form
	\begin{equation}\label{eq7*} \dfrac{d^{2}\mathbf{\Psi}}{dX^{2}}-\dfrac{2}{\hbar^{2}}X^{2(\frac{1}{n}-1)}V[X]\mathbf{\Psi}(X)=0.
	\end{equation} The problem of non-unitary evolution can be resolved by proper operator ordering in the first term of the Hamiltonian. \\If one considers the following operator ordering 
	\begin{equation}
		\tilde{\mathcal{H}}=-\dfrac{\hbar^{2}}{2} X^{(1-\frac{1}{n})}\dfrac{d}{dX} X^{(1-\frac{1}{n})}\dfrac{d}{dX}+V[X],
	\end{equation} then by choosing $v=nX^{\frac{1}{n}}$ the WD equation is transformed as 
	\begin{equation}\label{eq7}
		\left(-\dfrac{\hbar^{2}}{2} \dfrac{d^{2}}{dv^{2}}+V[v]\right)\mathbf{\Psi}(v)=0,
	\end{equation} with symmetric norm as $|\mathbf{\Psi}|^{2}=\int_{0}^{\infty} dv \mathbf{\Psi}^{*} \mathbf{\Psi}$, provided the integral exists and finite. \\ \\
	Takeaway from this quantization is as follows:
	\begin{itemize} 
		\item The WD equation (\ref{eq7}) can be interpreted as time-independent Schrödinger equation of a point particle of unit mass moving along $v$ direction in a potential field $V(v)$ and it has zero eigen  value of the Hamiltonian and the wave function of the universe is identified as the energy eigen function. 
		\item 
		$V [v]$ is the potential corresponding to
		the dynamical system representing the congruence and has to be constructed using the
		gravitational field equations. In case of homogeneous cosmology $V[v]=V(v)$. For inhomogeneous cosmologies, the method will not work,
		as $V$ will remain a functional. 
		\item For any modified gravity theory constructed in the background of homogeneous and isotropic space-time one may find $V$. Further if one can solve the WD equation and find $|\mathbf{\Psi}|^{2}$  then it can be used as a quantity proportional to the probability measure on the minisuperspace. If $|\mathbf{\Psi}|^{2}=0$ at classical singularity then singularity is avoided otherwise the singularity still persists in the quantum description. \item One may note that the existence (or non existence) of singularity is not a generic one, it depends on the gravity theory under consideration through the convergence scalar $R_{c}$ via the classical potential $V$. 
		\item This quantization is expected to find application  in the investigation of the
		singularities in the quantum regime for a collapse of homogeneous systems, such as the
		Datt-Oppenheimer-Snyder collapse \cite{Choudhury:2019snb}.
		\item Thus the  quantization of the geometric flow of a congruence of  classical geodesic gives possibilities of avoidance of singularity in homogeneous cosmologies.
		\item This WD quantization technique has been employed in \cite{Chakraborty:2023yyz} where the possible avoidance of singularity has been shown in some physically motivated $f(T)$ gravity models.
		\item One drawback of this quantization is that it is applicable only in case of homogeneous cosmologies but applies to both isotropic as well as anisotropic universe.
	\end{itemize}
	\subsection{Bohmian Trajectories}
		Unlike the previous section where we quantized the geodesic flow in this section we replace them by quantum Bohmian trajectories and find their explicit expressions. To do so, one may choose the ansatz for the wave function as  \cite{Chakraborty:2001za}
	\begin{equation}\label{eq90*}
		\mathbf{\Psi}(X)= A(X)~\exp\left(\frac{i}{\hbar}~S(X)\right).
	\end{equation}
	Using this ansatz into the WD - equation (\ref{eq7*}) one gets the Hamilton-Jacobi equation as 
	\begin{equation}
		\dfrac{-1}{2~X^{2(\frac{1}{n}-1)}}~\left(\dfrac{dS}{dX}\right)^{2} + V_{Q} + V(X)=0,
	\end{equation}
	where $V_{Q}$, the quantum potential has the expression as
	\begin{equation}
		V_{Q}=\dfrac{1}{2 A(X)~X^{2(\frac{1}{n}-1)}} \dfrac{d^{2}A(X)}{dX^{2}}.
	\end{equation}  Thus the Hamilton-Jacobi function $S$ is given by 
	\begin{equation}
		S=s_{0} \pm \int\left(\dfrac{1}{A(X)}\dfrac{d^{2}A(X)}{dX^{2}}+ 2x^{2(\frac{1}{n}-1)}\right)^{\frac{1}{2}}~dX,\label{eq189}
	\end{equation} where $s_{0}$ is the constant of integration.\\
	It may be noted that the trajectories $X(t)$ due to causal interpretation should be real, independent of any observation and are classified by the above H-J equation. In fact, the quantum trajectories i.e the Bohmian trajectories are first order differential equations characterized by the equivalence of the usual definition of momentum with that from the Hamilton-Jacobi function $S$ as 
	\begin{equation}
		\dfrac{dS(X)}{dX}=-2X^{2(\frac{1}{n}-1)}\dfrac{dX}{dt}
	\end{equation}or,
	\begin{equation}\label{eq96}
		2~\int \dfrac{X^{2(\frac{1}{n}-1)}~dX}{\left(\dfrac{1}{A(X)}~\dfrac{d^{2}A(X)}{dX^{2}}+2x^{2(\frac{1}{n}-1)}\right)^{\frac{1}{2}}}=\mp (t-t_{0}).
	\end{equation}
	For a universe
	described by an FLRW metric containing a distribution of perfect fluid having an equation
	of state $p = \epsilon \rho$, where $p$, $\rho$ are the pressure and density of the fluid with $\epsilon$ being a
	constant, one has solution for $a$ as,
	\begin{equation}
		a(t)=a_{0}(t-t_{0})^{K},
	\end{equation} where $a_{0},~t_{0},~K$ are constants.
	Now we consider the following cases:\\
	\textbf{Case-A}
	$A(X)=A_{0}$, a constant then from (\ref{eq189}) one has $
	S=s_{0}\pm \sqrt{2}~n~x^{\frac{1}{n}}
	$ and the quantum trajectory is described as,
	\begin{equation}
		\sqrt{2}~n~X^{\frac{1}{n}}=\pm (t-t_{0}).
	\end{equation}
	Here the quantum potential ($V_{Q}$) is zero and the H-J equation coincides with the classical one. Thus Bohmian trajectory corresponds to classical power law form of expansion and it can't avoid the initial big-bang singularity. \\ \\
	\textbf{Case-B} $A(X)=X^{L}$, $L\in(0,1)$ and substituting this in equation (\ref{eq96}) one gets the quantum trajectories as,
	\begin{equation}
		X(t)=\left[\frac{(t-t_{0})^{2}}{2n^{2}}-\frac{L(L-1)}{2}\right]^{\frac{n}{2}}.
	\end{equation}
	In this choice the quantum potential has non zero value and the Bohmian trajectories can avoid the initial classical singularity.\\ \\
	\textbf{Case-C} $A(X)=A_{0}\exp(-\alpha X),~\alpha\neq0$.
	The quantum trajectory is described by
	\begin{equation}
		\dfrac{6}{\alpha X^{\frac{1}{3}}}2^{F_{1}}\left(\dfrac{1}{4},\dfrac{1}{2},\dfrac{5}{4},-\dfrac{2}{\alpha^{2}X^{\frac{4}{3}}}\right)=(t-t_{0}).
	\end{equation}  where $2^{F1}$ is the Gauss-Hypergeometric function. Clearly the quantum trajectory is a one parameter family of curves which never pass through classical singularity.
	\section{Discussions and Future scope}
	To conclude, this review aims at a revisit to the classical and quantum consequences of the RE and singularity analysis in GR as well as in Modified theories of gravity. To mitigate the cosmological as well as Black-Hole singularities, different techniques have been adopted along with further scope of application. The wide range of applicability of the RE may be attributed to the fact that the equations encode geometric statements about flows. In physics, flows appear in diverse contexts and so is the applicability of the RE, as the later is nothing but an evolution for the geodesic flow. Understanding the singularity theorems classically and restating them quantum mechanically can be a good piece of future work. That means, it is exciting to find how the classical singularity theorems will look like if its quantum analogue is being studied? \\ 
	
	Although the quantum corrections to RE has been studied extensively as a tool to avoid the Black-Hole singularity, more applications can be studied in greater details. Since, RE is associated with geodesic flow, it can also be made analogous to geometric flows particularly the Ricci flow so that if the analogy is established then properties and theorems of Ricci flow can be applied for more exciting results to come up. Since the kinematic quantities that appear on the r.h.s of the RE are derivatives of a vector field. This hints that wherever there is a vector field, we can deduce some analogous form of the RE.\\
	
	On the other hand, there are a plethora of gravity theories where one may formulate the RE and modified CC for examining singularity free nature of underlying theory and background space-time. Further the application of RE and more importantly the singularity theorems in studying collapse of a star and its stability is of greater importance since these include energy conditions. Finally Bohmian trajectories can be made more elaborative in the sense that generalization of the Bohmian formalism may come up with interesting comments on the nature of the trajectories near the classical Big-Bang singularity. These are some possible scope of future work. We end this discussion, stating clearly the fact that there are diverse fields where RE can be employed, of which we have touched only a few. 
	\section*{Acknowledgment}
	The authors thank the respected editor Prof.(Dr.) Salvatore Capozziello and Prof.(Dr.) P.K.Sahoo for giving them the opportunity to write this review under the special issue titled ``From Cosmology
	to Quantum Information - Exploring the Frontiers of Geometric Methods in
	Physics''. M.C thanks University Grant Commission (UGC) for providing the Junior Research Fellowship (ID:211610035684/JOINT CSIR-UGC NET JUNE-2021). S.C. thanks FIST program of DST, Department of Mathematics, JU(SR/FST/MS-II/2021/101(C)).
	
\section{Appendix}
We consider the ultrastatic Morris-Thorne wormhole described by the line element
\begin{equation}
		ds^{2}=- dt^{2}+\left(1-\dfrac{b(r)}{r}\right)^{-1}dr^{2}+r^{2}d\Omega_2^{2}
\end{equation} Without loss of generality we consider $\theta=\dfrac{\pi}{2}$ and the components of the time-like velocity vector field $u^{\mu}$ is given by
\begin{equation}
	\dot{t}=-E ,~ \dot{\phi}=\dfrac{h}{r^2}  ,~ \dot{r}=\sqrt{\left(1-\frac{b}{r}\right)\left(E^2-1-\frac{h^2}{r^2}\right)},~\dot{\theta}=0
\end{equation}
where $E$ and $h$ are identified as the conserved energy and angular momentum of the time-like particle (per unit mass). The explicit expression for $R_{\mu\nu}u^{\mu}u^{\nu}$ is as follows : 
\begin{equation}
\tilde{R}=R_{\mu\nu}u^{\mu}u^{\nu}	=-\frac{h^2}{r^4}+\frac{b'(r)}{r^2}\left(E^2-1-\frac{h^2}{2r^2}\right)-\frac{b(r)}{r^3}\left(E^2-1-\frac{3h^2}{2r^2}\right)
\end{equation}
where $R_{\mu\nu}$ is the Ricci tensor projected along the congruence of geodesics and it has been evaluated from the metric. For realistic $\dot{r}$, we need \begin{equation}
	(E^{2}-1)>\dfrac{h^{2}}{r^{2}}\label{eq213}
\end{equation}
 From the metric we have \begin{equation}
 \left(1-\dfrac{b(r)}{r}\right)>0.\label{eq214}
 \end{equation} The flairing out condition gives
\begin{equation}
	\dfrac{-rb'+b}{b^{2}}>0\nonumber
\end{equation} which upon simplification yields,
\begin{equation}
	\dfrac{b(r)}{r^{3}}-\dfrac{b'(r)}{r^{2}}>0 \label{eq215}
\end{equation}
After some algebraic manipulation, one can write $\tilde{R}$ as
\begin{equation}
	\tilde{R}=R_{1}+R_{2}+R_{3}
	\end{equation} where,
\begin{eqnarray}
	R_{1}=\left(E^{2}-1-\dfrac{h^{2}}{r^{2}}\right)\left(\dfrac{b'(r)}{r^{2}}-\dfrac{b(r)}{r^{3}}\right)\\ R_{2}=\dfrac{1}{2}\dfrac{h^{2}}{r^{2}}\left(\dfrac{b'(r)}{r^{2}}-\dfrac{b(r)}{r^{3}}\right)\\
	R_{3}=\dfrac{h^{2}}{r^{4}}\left(\dfrac{b(r)}{r}-1\right)
\end{eqnarray}
Now equations (\ref{eq213}), (\ref{eq214}) and (\ref{eq215}) show that $R_{1}, R_{2}, R_{3}$ all are negative and hence $\tilde{R}<0$. Thus CC is violated inside the wormhole with time-like test particle just like in the case of null geodesics.
\end{document}